\documentclass[journal=jacsat,manuscript=article,layout=twocolumn]{achemso}

\usepackage[version=3]{mhchem} 
\usepackage{cleveref}
\crefname{equation}{Eq.}{Eqs.} 
\usepackage{titlesec}
\usepackage{tabularx}
\usepackage{booktabs}

\newcolumntype{Y}{>{\raggedright\arraybackslash}X}

\author{Hodaka Mori}
\email{hodakamori@preferred.jp}
\affiliation[Preferred Networks]{Preferred Networks, Inc., Tokyo 100-0004, Japan}

\author{Shunsuke Tonogai}
\affiliation[Preferred Networks]{Preferred Networks, Inc., Tokyo 100-0004, Japan}

\author{Yu Miyazaki}
\affiliation[Preferred Networks]{Preferred Networks, Inc., Tokyo 100-0004, Japan}

\author{Akihide Hayashi}
\affiliation[Preferred Networks]{Preferred Networks, Inc., Tokyo 100-0004, Japan}

\author{Masayoshi Takayanagi}
\affiliation{
Data Science and AI Innovation Research Promotion Center, 
Shiga University, Hikone 522-8522, Japan
}

\title[An \textsf{achemso} demo]
  {Ready-to-Use Polymerization Simulations Combining Universal Machine Learning Interatomic Potential with Time-Dependent Bond Boosting for Polymer and Interface Design}

\begin{document}

\begin{tocentry}
  \centering
  \includegraphics[
    width=\linewidth,height=4.5cm,
    keepaspectratio,clip,
  ]{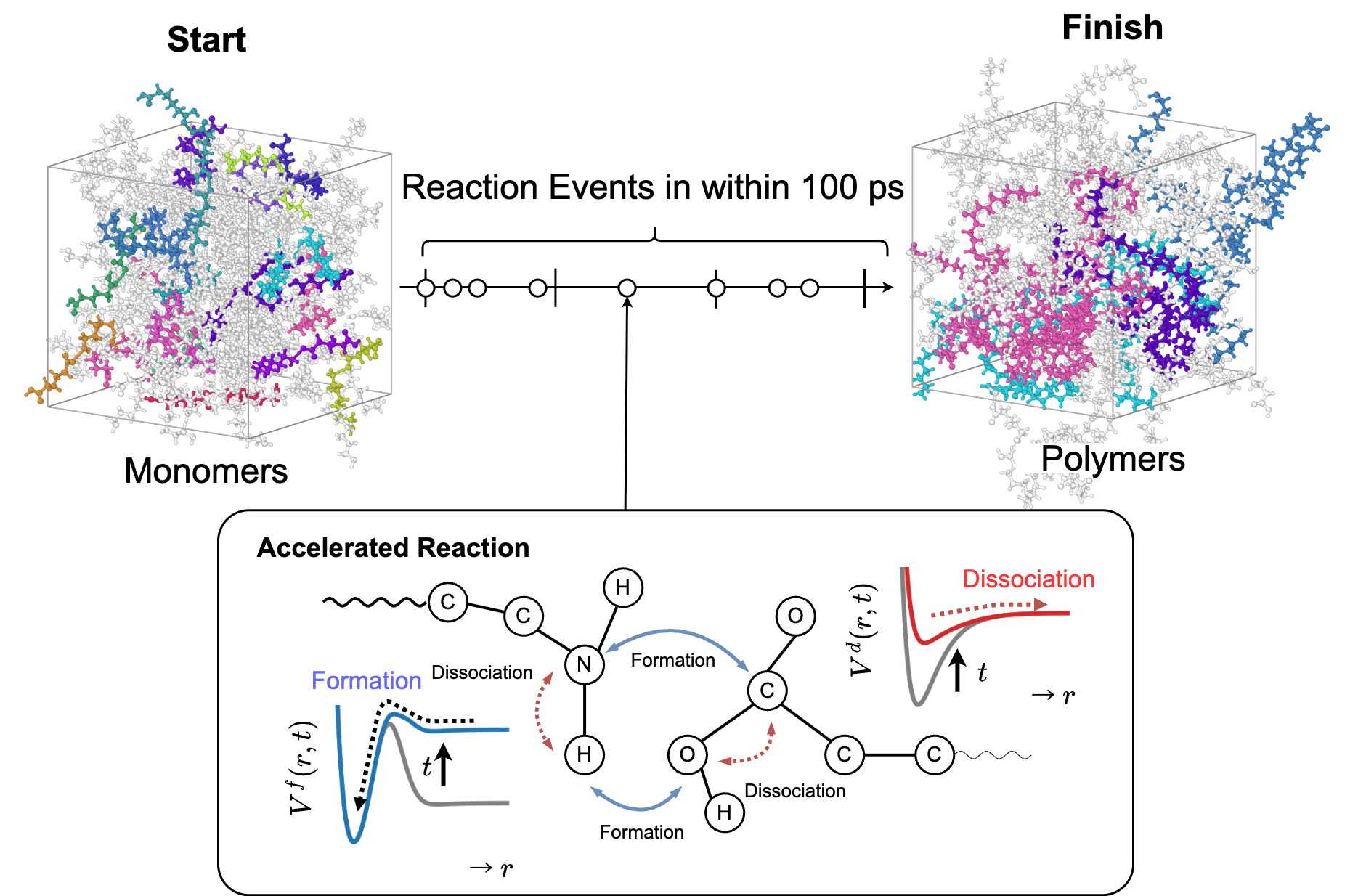}
\end{tocentry}

\begin{abstract}
Although polymerization and curing reactions govern the performance of advanced materials, their simulation remains challenging owing to the need for accurate, transferable potentials and rarity of chemical events. Conventional reactive force fields such as ReaxFF require system-specific parametrization, while universal machine learning interatomic potentials (uMLIPs) exhibit limited sampling efficiency. This paper introduces a novel simulation framework integrating a uMLIP with a time-dependent bond-boost scheme. The bias potential increases monotonically with time, and the use of a unified parameter set across reaction classes enables consistent acceleration without system-specific tuning. For radical polymerization of vinyl monomers, the proposed framework reproduces characteristic trends, such as linear molecular-weight growth with conversion, initiator-concentration scaling, and relative monomer reactivity trends. For step-growth polycondensation of nylon-6,6, it captures the characteristic sharp increase in molecular weight at high conversion rates, consistent with experimental behavior. For epoxy curing at a copper substrate, it reveals interfacial ring-opening and cross-linking events, consistent with spectroscopic evidence of Cu—O--C bond formation. Overall, coupling uMLIPs with time-dependent bond boost enables practical and transferable simulations of polymerization and curing processes. The proposed framework reliably resolves mechanistic pathways and relative reactivity, offering molecular-level insights into polymer growth and interfacial adhesion.
\end{abstract}

\section{Introduction}
Polymeric materials play an indispensable role across a wide range of industrial applications, including electronics, medical devices, energy storage, and structural components\cite{McCrum1997-ge,Ebewele2000-fl,Young2011-oz,Wypych2022-lf}. Their performance strongly depends on the molecular structures formed by microscopic chemical processes such as polymerization and cross-linking, highlighting the importance of understanding structural changes and reaction mechanisms at the molecular level\cite{Rao2022-jj, Bezik2023-ts}. Thus, to design high-performance materials and develop novel polymeric systems, the corresponding reaction pathways, intermediates, and transition states must be comprehensively explored.

Experimental techniques such as time-resolved spectroscopy and electron microscopy have been used to analyze polymer reactions\cite{Pramanik2014-uy, Beuermann2002-ps}. However, direct observation of reaction intermediates and detailed mechanisms at the atomic level remains challenging\cite{Gartner2019-lq, Joshi2021-kh, Krishna2021-ku, Liu2021-og}. Molecular simulations serve as powerful complementary tools that allow the identification of reaction intermediates and transition states that are difficult to capture experimentally\cite{Arash2017-oi, Mori2018-be, Odegard2021-nc, Yamaguchi2022-je}. By providing atomistic insights, these simulations offer valuable guidelines for the rational design of advanced materials.

Two methods are widely used for simulating cross-linking and polymerization reactions: pseudoreaction models and reactive potentials. Pseudoreaction models stochastically rearrange chemical bonds when specified distance or angular criteria are met\cite{Okabe2013-lk,nagaoka2013hybrid,suzuki2017transformation,Gissinger2017-wh,Gissinger2024-nx}. Although these models can be combined with classical molecular dynamics (MD) to handle large systems, they necessitate complex parameter tuning, including the definition of activation energies and frequency factors in advance. Quantum chemical approaches such as global reaction route mapping can accurately calculate activation energies. However, their computational cost remains prohibitive for large or structurally diverse condensed-phase systems\cite{Rao2022-jj, Xi2024-hl,Xi2024-lq}.

A well-known reactive force field approach is ReaxFF, which describes the formation and breaking of chemical bonds during MD simulations\cite{Van_Duin2001-ag}. However, many chemical reactions in polymerization processes are rare events, meaning that straightforward nanosecond-scale MD simulations often fail to capture sufficient reaction progress. This problem becomes particularly pronounced in polymerization, which involves a large number of sequential reactions, necessitating unrealistically long simulation times\cite{brandrup1999polymer}. To address this, acceleration techniques such as bond-boost MD have been developed\cite{Vashisth2018-kb, Dasgupta2020-ll}. However, the creation and parametrization of specialized force fields for each target system require substantial expertise and effort. Moreover, the bond-boost method involves hyperparameters that, if set improperly, may compromise the reliability of the observed reactions. Another widely used approach for enhancing the sampling of rare events is metadynamics, which applies history-dependent bias potentials to efficiently explore free-energy landscapes \cite{Laio2002-kx,Invernizzi2020-dg}. However, by design, metadynamics promotes both forward and backward reactions, making it less suitable for directly accelerating one-way polymerization processes.

Given these limitations, there exists a strong demand for simulation methods that are simple, accurate, and broadly applicable to diverse polymerization processes. Recently, universal machine learning interatomic potentials (uMLIPs) trained on density functional theory data have attracted increasing attention in this domain.\cite{Kaser2023-la, Duignan2024-vm, Martin-Barrios2024-qa} In particular, uMLIPs such as PreFerred Potential (PFP), M3GNet, CHGNet, and MACE have demonstrated remarkable generalization across various atomic species and structures,\cite{Batatia2022-eh, Batatia2025-zd, Takamoto2022-mj, Chen2022-en, Deng2023-ms} enabling their application to catalysis, metal--organic frameworks, and solid--liquid interfaces.\cite{Tayfuroglu2022-ei, Hisama2024-jc, Hisama2024-ey, Lin2025-sw} The incorporation of these uMLIPs largely eliminates the need to construct system-specific force fields, thereby broadening the scope of atomistic simulations.

Considering these aspects, this paper introduces a time-dependent reaction acceleration strategy that combines the broad transferability of a uMLIP with efficient sampling. This approach enables polymerization simulations that are both accurate and straightforward to implement, even for chemically complex systems. To validate the method, we first benchmark the radical polymerization of vinyl monomers, focusing on kinetics and relative reactivity. We then demonstrate its broad applicability through step-growth polycondensation of nylon-6,6 and the curing of an epoxy resin at a copper oxide (CuO) interface, a system known to be particularly challenging to treat using conventional reactive force fields such as ReaxFF. Collectively, these applications highlight the potential of the proposed framework to provide accurate and convenient polymerization simulations across diverse chemical processes.

\section{Computational Details}
\subsection{Time-dependent Bond-Boost Potential}
\label{sec:tdbondboost}

To accelerate reactions, we use a modified version of the bond-boost method originally proposed by Vashisth et al.~\cite{Vashisth2018-kb}. 
The original bond-boost potential is expressed as
\begin{equation}
V(r)=f_1 \Bigl\{1-\exp\bigl[-f_2(r-r_0)^2\bigr]\Bigr\}.
\end{equation}
Here, $f_{1}$, $f_{2}$, $r_{0}$, and $r$ represent the depth of the boost potential, its range, the target distance, and the interatomic distance, respectively. 
Appropriate values for \(f_{1}\), \(f_{2}\), and \(r_{0}\) must be specified prior to the simulation. 
Vashisth et al.~\cite{Vashisth2018-kb} and Dasgupta et al.~\cite{Dasgupta2020-ll} examined multiple parameter sets and adopted the acceleration conditions that best reproduced the activation energies obtained using ReaxFF.
Typically, determining $f_{1}$ requires estimating the activation energy in advance and then fine-tuning the parameters through trial and error. 
Furthermore, for bond cleavage, an appropriate distance \(r_{0}\) must be defined as the threshold for bond breaking. 
As the number of target reactions increases, such preliminary parameter studies become increasingly cumbersome, complicating the overall simulation setup.

To address this, in this work, we introduce a time-dependent bond-boost (TDBB) method as a modification of the original approach. 
The key innovation is that the bias potential increases with time, as shown in \Cref{eq:Vf,eq:Vd,eq:r0,eq:f1}. 
Here, \(V^{f}(r,t)\) and \(V^{d}(r,t)\) represent the bias potentials applied to promote \textbf{bond formation} and \textbf{bond dissociation}, respectively:

\begin{equation}
V^{f}(r,t) = f_1(t) \Bigl\{1-\exp\bigl[-f_2(r-r_0)^2\bigr]\Bigr\}, 
\label{eq:Vf}
\end{equation}
\begin{equation}
V^{d}(r,t) = f_1(t) \exp\bigl(-f_2 r^2\bigr), 
\label{eq:Vd}
\end{equation}
\begin{equation}
r_0 = \lambda \sum_{a \in \{\text{pair}\}} r_a^{\mathrm{vdw}},
\label{eq:r0}
\end{equation}
\begin{equation}
f_1(t)=
\begin{cases}
\gamma t, & f_1(t) < f_1^{\mathrm{max}},\\
f_1^{\mathrm{max}}, & f_1(t) \geq f_1^{\mathrm{max}},
\end{cases}
\label{eq:f1}
\end{equation}

where $\gamma$ is the boost factor, $r_a^{\mathrm{vdw}}$ is the van der Waals radius of atom $a$ (summed over the two atoms forming the reactive pair), $\lambda$ is a scale factor defining the bonding threshold, and $f_1^{\mathrm{max}}$ is the maximum boost amplitude. 
Because $f_1(t)$ increases with time, reactions are gradually promoted as the simulation proceeds. 
Moreover, applying a single set of boost parameters to all possible reactions ensures consistent acceleration and the capturing of relative reactivities without system-specific parameterization. 
The overall acceleration level can be tuned by $\gamma$, with larger values yielding higher reaction rates.

Notably, we employ distinct functional forms for bond formation, $V^{f}(r_{ij})$, and bond breaking, $V^{d}(r_{ij})$, to ensure that once the product or dissociated state is reached, the bias potential vanishes, and normal MD resumes. 
In the bond-formation case, the potential drives two atoms to approximately 60\% of the sum of their van der Waals radii. Once that distance is reached, the potential effectively becomes zero. 
In the bond-breaking case, once the bond extends beyond a critical length, the boost potential similarly disappears. 
This design eliminates the need to define separate parameters for each reaction, thereby simplifying the simulation setup.

According to \Cref{eq:Vf,eq:Vd,eq:r0,eq:f1}, the parameters $f_1$, $f_2$, and $r_0$ correspond closely to those in the original bond-boost method. 
Although these parameters can be fine-tuned if desired, $f_1^{\mathrm{max}}$ primarily serves to cap the boost strength and prevent excessively large acceleration. 
Similarly, $f_2$ controls how the boost varies with distance: Smaller values concentrate the effect at very short range (strong near contact), whereas larger values extend the influence over longer distances. 
By setting $r_{0,\mathrm{create}}$ to the sum of the van der Waals radii of the involved atoms, this parameter becomes independent of the reaction pathway. 
In practice, the same settings have been successfully applied to multiple reactions.

The basic functional form follows the original bond-boost design, ensuring that once a reaction occurs, the boost potential decays to zero and the system behaves as in a standard MD simulation without acceleration. 
An additional advantage of this straightforward Morse-like potential is that it consistently exerts force along the direction of the intended reaction coordinate. 
In contrast, adaptive bias methods such as metadynamics can, in some cases, drive the system in the opposite direction of the desired pathway. 
Although adding a high-wall restraint can prevent the bias from diverging, such methods typically fill energy minima before effectively accelerating the system toward the reaction coordinate, potentially delaying the onset of meaningful acceleration. 
Moreover, in metadynamics, the bias can continue to accumulate even after a reaction has occurred, necessitating additional steps (e.g., detecting the occurrence of the reaction) to halt bias deposition.

Lastly, the proposed method does not allow time rescaling, unlike conventional hyperdynamics~\cite{Voter1997-tf, Miron2003-ed, Bal2015-kj}. 
According to transition-state theory, time rescaling in biased MD is valid only if (1) the added potential does not exceed the activation energy and (2) no potential is applied at the dividing surface. 
In contrast, in the proposed approach, the bias grows monotonically with time and can surpass the activation barrier regardless of the potential energy landscape. 
Thus, these conditions are generally not satisfied. 
Notably, while this method simplifies parameter setup and effectively drives chemical reactions, the time evolution of the trajectory no longer strictly corresponds to physical time once the bias becomes substantial.

\subsection{Polymerization Procedure}
\begin{figure*}
    \centering
    \includegraphics[width=1.0\linewidth]{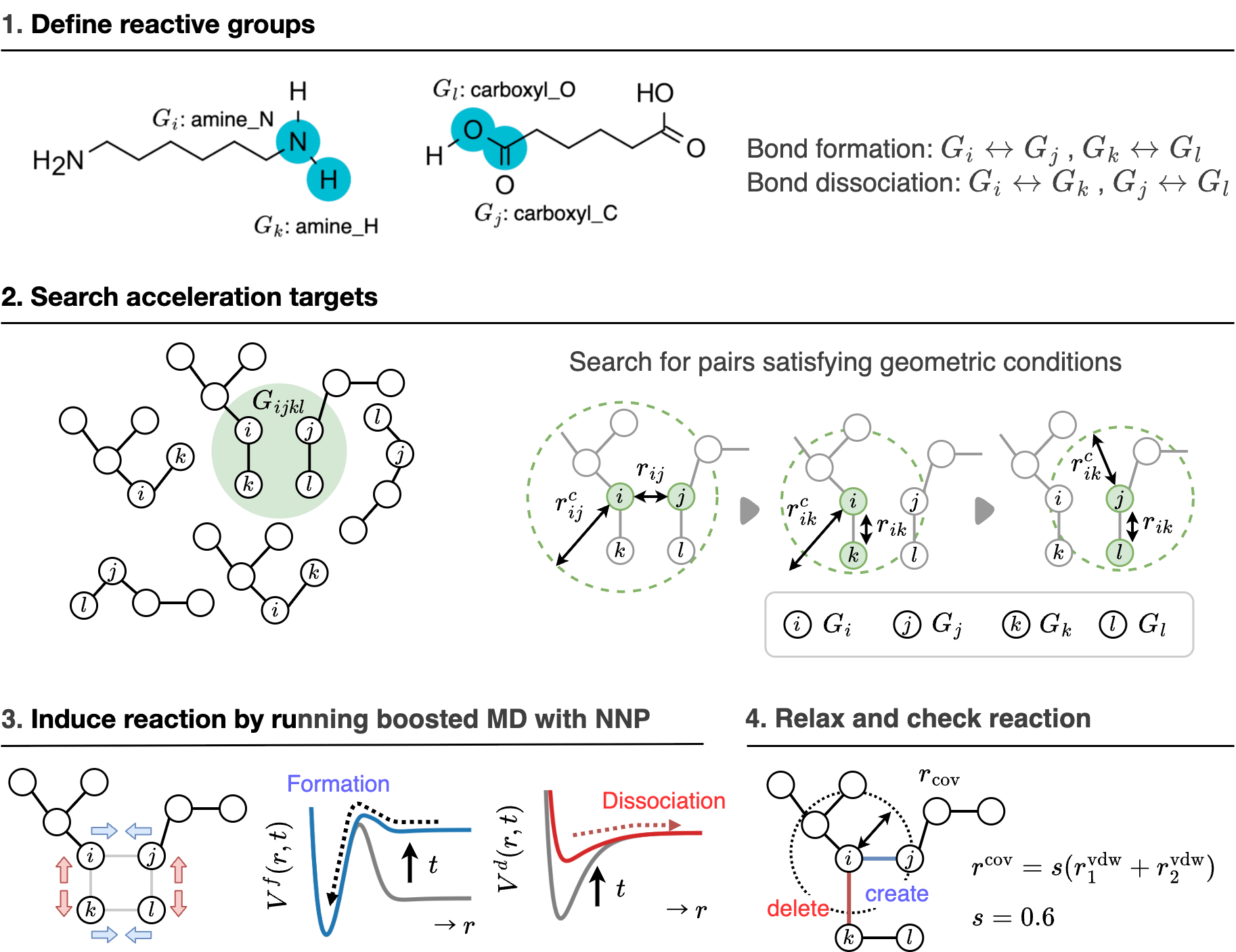}
    \caption{
    Schematic of the proposed simulation workflow, including 
    four main stages:
    (1) definition of reactive atomic groups, where potential reactive species are identified from chemical functionalities; 
    (2) selection and acceleration of reaction-targeted groups, where candidate atomic combinations satisfying geometric and chemical criteria are chosen and subjected to time-dependent bias potentials; 
    (3) reaction induction and structural relaxation, in which accelerated molecular dynamics simulations promote bond rearrangements followed by unbiased relaxation; and 
    (4) reaction evaluation and group update, where the bonding states are analyzed and reactive groups are updated for subsequent iterations.
    }
    \label{fig:algo}
\end{figure*}

The proposed simulation method involves four main steps (Figure~\ref{fig:algo}): 
(1) definition of reactive atomic groups, 
(2) selection and acceleration of reaction-targeted groups, 
(3) reaction induction and structural relaxation, and 
(4) evaluation of reactions and updating of atomic groups.

\textbf{1. Definition of Reactive Atomic Groups:} 
In the initial stage, reactive atomic groups are identified and defined as sets \(G_I\), \(G_J\), \(G_K\), and \(G_L\), each corresponding to a specific group (e.g., oxygen atoms in epoxy groups, nitrogen atoms in amine groups, or \(\alpha\)-carbon atoms in vinyl groups for radical polymerization). 
These sets are defined as
\begin{equation}
G_X = \{a \mid a \in X\},\quad X = I, J, K, L,
\end{equation}
with $a$ representing an individual atom belonging to the group. 
Chemical reactions are modeled as rearrangements of atomic bonds, represented by bond formation and breakage.

\textbf{2. Selection and Acceleration of Reaction-Targeted Groups:} 
To efficiently identify possible reaction candidates, a two-stage selection process is employed. 
First, neighboring atomic groups are preselected based on geometric proximity to identify potential reacting partners. 
Subsequently, based on reaction rules, candidate groups are formed by selecting tuples $(i,j,k,l)$ such that 
$i \in G_I$, $j \in G_J$, $k \in G_K$, and $l \in G_L$, with interatomic distances constrained to the window
$r_{ab}^{\min} \le r_{ab} \le r_{ab}^{\max}$. 
Defining the set of target pairs as $\mathcal P = \{(i,j),(i,k),(j,l)\}$, the candidate set is expressed as
\begin{equation}
\begin{aligned}
G_{ijkl} = \{(i,j,k,l) \mid &\, i \in G_I,\, j \in G_J, \\
&\, k \in G_K,\, l \in G_L,\\
&\, r_{ab}\in [r_{ab}^{\min}, r_{ab}^{\max}] \\
&\, \forall (a,b) \in \mathcal P\}.
\end{aligned}
\end{equation}

Moreover, we allow multiple reaction types to coexist. 
For each reaction type $r$, a candidate set $G_{ijkl}^{(r)}$ is prepared with its own group assignments $(G_I,G_J,G_K,G_L)$, 
target-pair set $\mathcal{P}^{(r)}$, and distance windows $\bigl[r_{ab}^{\min},r_{ab}^{\max}\bigr]^{(r)}$. 
This setup enables systems in which different reactions proceed simultaneously within a single simulation box (e.g., 
Dehydration--condensation and radical polymerization occurring in parallel). 
Unless otherwise noted, candidates from all reaction types are pooled and subjected to the same nonoverlapping selection and biasing procedure described below.

As candidate groups may share atoms (e.g., \((0,1,2,3)\) and \((0,4,5,6)\)), 
a nonoverlapping selection is performed by computing the score 
$d_{ijkl} = r_{ij} + r_{ik} + r_{jl}$ for each group, sorting the obtained scores in ascending order, 
and sequentially selecting groups while skipping those with overlapping atoms. 
This yields the refined set $G_{ijkl}^{\mathrm{acc}}$. 
For each group in $G_{ijkl}^{\mathrm{acc}}$, the set of reactive atomic pairs $P$ is then defined, 
consisting of one or more pairs selected from $(i,j)$, $(i,k)$, $(j,l)$, and $(k,l)$ that participate in bond formation or cleavage. 
A time-dependent bias potential is subsequently applied to each pair in $P$:
\begin{equation}
\begin{aligned}
\Delta V(t) &= 
\sum_{(i,j,k,l) \in G_{ijkl}^{\mathrm{acc}}} 
\sum_{p \in P} \Bigl[ \\
&\quad f_{p}\, V^{f}(r_{p},t) \\
&\quad + \bigl(1 - f_{p}\bigr)\, V^{d}(r_{p},t) \Bigr],
\end{aligned}
\end{equation}
where $f_{p} = 1$ denotes bond formation, $f_{p} = 0$ denotes bond breakage, 
and $r_{p}$ is the interatomic distance of pair $p$.

\textbf{3. Reaction Induction and Structural Relaxation:} 
The system is subjected to an accelerated MD simulation under the applied bias $\Delta V(t)$ for either a predetermined period or until a reaction event is observed. 
A reaction event is defined to occur when the interatomic distances of a specific set of pairs (\(ij\), \(ik\), and \(jl\)) simultaneously satisfy the prescribed bonding conditions. 
For each pair, the bonding state is determined based on the van der Waals criterion: Two atoms are considered bonded if their separation falls below 60\% of the sum of their van der Waals radii. 
If $f_{p}=1$, the event is triggered when these pairs newly satisfy the bonding condition (bond formation). If $f_{p}=0$, the event is triggered when the pairs cease to satisfy the condition (bond dissociation). 
Once such an event is detected, or the maximum simulation time is reached, the acceleration phase terminates. 
Subsequently, a standard MD simulation (without the bias) is performed to allow structural relaxation and equilibration of the resulting configurations.

\textbf{4. Reaction Evaluation and Group Update:} 
After structural relaxation, the system is analyzed to compare bonding states before and after the biased simulation. 
Atoms that have participated in new bond formations or bond cleavages are marked as ``reacted'' and removed from the reactive set. 
In contrast, atoms that experienced the bias but did not undergo any bond changes remain in the reactive set for subsequent iterations. 
If the resulting species remain reactive (e.g., in epoxy--amine curing or vinyl radical polymerization), these atoms are reassigned to Group~$X$ and considered in subsequent iterations. 
Repeating these steps iteratively drives the overall polymerization process.

\subsection{Comparison with existing acceleration methods}

To highlight the novelty and advantages of the proposed framework, 
Table~\ref{tab:methods_comparison} summarizes the key features of widely used 
rare-event acceleration techniques. 
Conventional bond boost requires system-specific parameter tuning based on 
activation barriers, whereas hyperdynamics and metadynamics depend on carefully 
chosen collective variables and typically accelerate both forward and backward 
events indiscriminately. 
In particular, even infrequent metadynamics---the only variant that, in principle, 
allows approximate recovery of physical timescales---is intrinsically bidirectional: 
Gaussian hills are deposited regardless of whether the system proceeds forward or backward, 
which makes the method inefficient and physically inconsistent for inherently 
unidirectional processes such as radical polymerization. 
In contrast, bond-boost schemes apply a forward-only bias that selectively promotes 
bond formation, aligning with the irreversible nature of 
polymer growth. 
Our TDBB variant retains this property while 
eliminating the need for prior parametrization, allowing a single parameter set 
to be used across reaction types. 
Combined with uMLIP such as PFP, this approach 
offers a unique balance of simplicity, accuracy, and transferability, and is 
particularly well suited for chemically complex and heterogeneous polymerization systems.

\newcolumntype{Y}{>{\raggedright\arraybackslash}X}

\begin{table*}[htbp]
  \centering
  \footnotesize
  \caption{Comparison of acceleration methods, focusing on bias form and setup requirements.}
  \label{tab:methods_comparison}
  \begin{tabularx}{\textwidth}{YYYYY}
    \toprule
    Method & Setup requirements & Bias function & Reaction directionality & Time rescaling \\
    \midrule
    Bond Boost \cite{vashisth2018accelerated} 
      & Reactive groups, bias height, target distance
      & Fixed bias 
      & Unidirectional 
      & No \\
    Metadynamics 
      & Collective variables (CV)
      & Adaptive hill bias
      & Bidirectional
      & No \textsuperscript{\ddag} \\
    CVHD 
      & CV and dividing surface 
      & Adaptive hill bias 
      & Bidirectional
      & Yes\textsuperscript{*} \\
    TDBB (this work) 
      & Reactive groups
      & Monotonic, time-growing bias 
      & Unidirectional 
      & No \\
    \bottomrule
  \end{tabularx}

  \vspace{2mm}
  \footnotesize
  \raggedright
  \textsuperscript{*}\,Hyperdynamics time rescaling is possible if conditions are met (zero bias at dividing surface, bias $<$ barrier, TST assumptions).\\
  \textsuperscript{\ddag}\, Except infrequent metadynamics.
\end{table*}

\section{Methods}
We investigated three representative systems: radical polymerization of vinyl monomers, 
step-growth polycondensation of nylon-6,6, and epoxy curing at a CuO interface. 
All systems were prepared through a two-stage equilibration procedure: initial relaxation with 
a classical force field, followed by refinement using the PFP uMLIP under 
the {\it NPT} ensemble. Detailed compositions and equilibration protocols are provided in the 
Supporting Information.

All simulations were conducted using PFP (version~v6.0.0) in the 
\texttt{CRYSTAL\_PLUS\_D3} mode. PFP is a uMLIP trained on Perdew—Burke--Ernzerhof reference 
data and augmented with the D3 dispersion correction, which improves the reproducibility of 
condensed-phase properties such as density. Owing to its transferability across 96 elements, 
PFP has been applied to a wide range of materials, including inorganic and organic 
compounds, polymers, and catalytic systems. 

Reactive acceleration MD was performed using the boost potential described in 
Section~\ref{sec:tdbondboost}. Unless otherwise noted, simulations employed a time step of 
0.25\,fs, alternating biased and unbiased dynamics every 2000 steps (500\,fs), and bias 
parameters of $f_{1}^{\mathrm{max}} = 250$\,kcal/mol for $V^f$, $f_{1}^{\mathrm{max}} = 125$\,kcal/mol for $V^d$, $f_2 = 10 ~\text{\AA}^{-2}$, and $\gamma = 1.0$. 
Reaction events were identified based on van der Waals radii as detailed in 
Section~\ref{sec:tdbondboost}. 
For each condition, three independent simulations with randomized initial configurations were performed, 
and the reported results represent averages over these runs. 
System-specific simulation parameters, including system size, composition, and equilibration protocol, 
are provided in the Supporting Information.

To evaluate the robustness of the acceleration scheme, we examined the sensitivity of the cumulative number of reactions to variations in $f_{1}^{\mathrm{max}}$, $f_{2}$, and $\gamma$, using styrene radical polymerization as a representative system (Fig.~S4). 
The influence of $f_{1}^{\mathrm{max}}$ was negligible once it exceeded a moderate threshold: extremely low values suppressed reactions within the time window, whereas sufficiently large values simply shifted the reaction to occur later within the same window, after which the reactive-pair list was updated. 
System behavior was also insensitive to variations in $f_{2}$. 
Although $f_{2}$ formally defines the spatial range of the bias, propagation is governed by near-contact events, and $f_{2}\approx 10$ was sufficient to ensure consistent behavior. 
In contrast, increasing $\gamma$ accelerated all reaction channels uniformly, acting as a global rate scaling factor. 
These analyses demonstrate that the default parameter set is broadly representative and does not distort relative reactivity trends.

All MD simulations were carried out using OpenMM~ver.~8.1.1 \cite{eastman2023openmm}. Forces 
and energies were evaluated using PFP, with GPU selection depending on the system size: an 
NVIDIA Tesla V100 (16\,GB memory) was used for medium-sized systems, whereas larger systems were 
computed on an NVIDIA A100 (80\,GB memory). A custom interface coupled OpenMM with PFP. The 
bias potential was implemented via OpenMM--Torch, enabling TorchScript-defined energy and 
force terms to be directly incorporated. Force derivatives were evaluated using PyTorch’s 
automatic differentiation, eliminating the need for manual implementation. The 
workflow of reactive acceleration MD was managed using an in-house Python script.

\section{Results and Discussion}
\label{sec:results}
\subsection{Radical polymerization of vinyl monomers}
The simulations were designed with two primary objectives:
(i) reproduce the kinetics of radical polymerization under solution conditions, and
(ii) reproduce the relative reactivity order of representative vinyl monomers under bulk conditions.
The monomer set included methyl acrylate, methyl methacrylate, styrene, vinyl acetate, diphenylethylene, and dimethyl itaconate.
Polymerizations were performed both in toluene solution and in the bulk phase.
Toluene was selected as a representative nonpolar solvent commonly used in experimental radical polymerizations, whereas bulk simulations enabled direct assessment of solvent-free reactivity.
Together, these two environments and the diverse monomer set provided a stringent benchmark for evaluating the predictive capability of the proposed simulation framework.

\begin{figure*}
    \centering
    \includegraphics[width=0.9\linewidth]{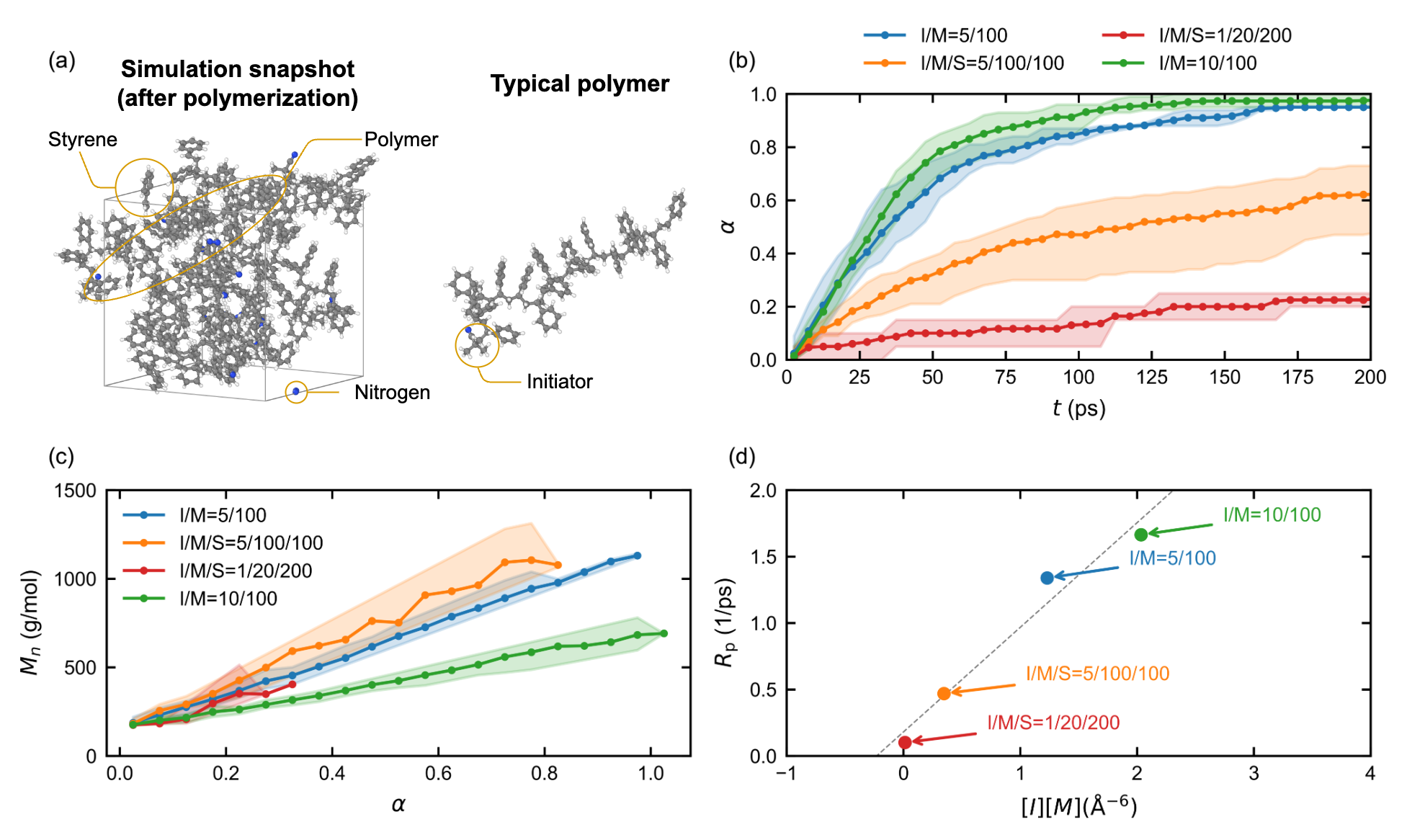}
    \caption{Structural and kinetic characteristics of styrene radical polymerization in toluene solvent.
(a) Snapshot after polymerization and a representative polymer chain.
(b) Time--conversion curves at different initiator-to-monomer (I/M) ratios.
The vertical axis $\alpha$ represents the monomer conversion, defined as 
$\alpha = 1 - [M]/[M]_0$, 
where $[M]_0$ and $[M]$ are the initial and instantaneous monomer concentrations, respectively.
(c) $M_\mathrm{n}$ as a function of conversion.
(d) Polymerization rate plotted against $[I][M]$. The horizontal axis represents the product of the initiator and monomer concentrations, $[I][M]$, converted from number densities to molar units (1~$\text{\AA}^{-6}$ = 2.76 × 10$^{-4}$ mol$^2$ L$^{-2}$).
Here, I, M, and S denote the initiator (AIBN), monomer (styrene), and solvent (toluene), respectively.}
    \label{fig:styrene}
\end{figure*}

Figure~\ref{fig:styrene} illustrates three key aspects of the styrene radical polymerization under varying concentrations: (i) time--conversion profiles, (ii) relationship between monomer conversion and number-average molecular weight ($M_n$), and (iii) dependence of the polymerization rate on the initiator-to-monomer ratio (I/M). As shown in Figure~\ref{fig:styrene} (b), systems with I/M ratios of 5/100 and 10/100 exhibit markedly higher conversions at earlier times compared to those with lower initiator loadings. This trend demonstrates that increasing the initiator and/or monomer concentrations accelerates polymerization, consistent with experimental observations and classical kinetic predictions. These results highlight the ability of the proposed uMLIP--accelerated MD framework to accurately capture the early-stage kinetics of radical polymerization.

The plots of conversion versus $M_n$ in Figure~\ref{fig:styrene} (c) exhibit a clear linear relationship across all simulated conditions. This linear trend strongly suggests that, in the absence of termination and chain-transfer reactions (see Figure.~S1), polymerization proceeds in a ``living'' manner, characterized by continuous chain growth without significant loss of active chain ends. Moreover, the slope of the conversion--molecular-weight curve for the I/M = 10/100 system is approximately half of that for the lower-initiator systems. This behavior is characteristic of living polymerization, where the number-average degree of polymerization is inversely proportional to the initiator concentration. This finding corroborates the fidelity of the proposed simulations in reproducing mechanistic features observed experimentally in controlled polymerization systems.

Figure~\ref{fig:styrene} (d) illustrates the dependence of the polymerization rate on the I/M ratio. 
In the absence of termination reactions, the polymerization rate $R_p$ is expressed as
\begin{equation}
R_p = -\frac{d[M]}{dt} = k_p[M][P^*]
\end{equation}

\begin{equation}
R_p \propto [M][I],
\end{equation}
where $[M]$ and $[I]$ denote the monomer and initiator concentrations, respectively; 
$[P^*]$ is the concentration of propagating radicals; and 
$k_p$ is the propagation rate constant. 
In this idealized case, all radicals generated from the initiator remain active, resulting in a linear dependence of the polymerization rate on initiator concentration. This is unlike the square-root dependence observed in conventional radical polymerizations.

The simulation results closely follow this theoretical proportionality, confirming that the proposed approach reproduces the linear dependence of the polymerization rate on initiator concentration and its proportionality to monomer concentration under the no-termination assumption. This agreement between simulation and theory not only validates the physical accuracy of the uMLIP but also underscores the robustness of the acceleration scheme in capturing fundamental kinetic behavior across a range of concentrations. The linear dependence of $R_p$ on initiator concentration is a hallmark of living radical polymerization, where active chain ends persist without termination.

Collectively, these results demonstrate that combining machine-learned potentials with accelerated dynamics can quantitatively reproduce key mechanistic and kinetic trends in radical polymerization. By accurately capturing the dependence of reaction rate and molecular weight evolution on initiator and monomer concentrations, the proposed framework provides a solid foundation for predictive simulations of polymerization and offers valuable insights into the molecular-level dynamics governing polymer growth.

\begin{figure*}
    \centering
    \includegraphics[width=0.9\linewidth]{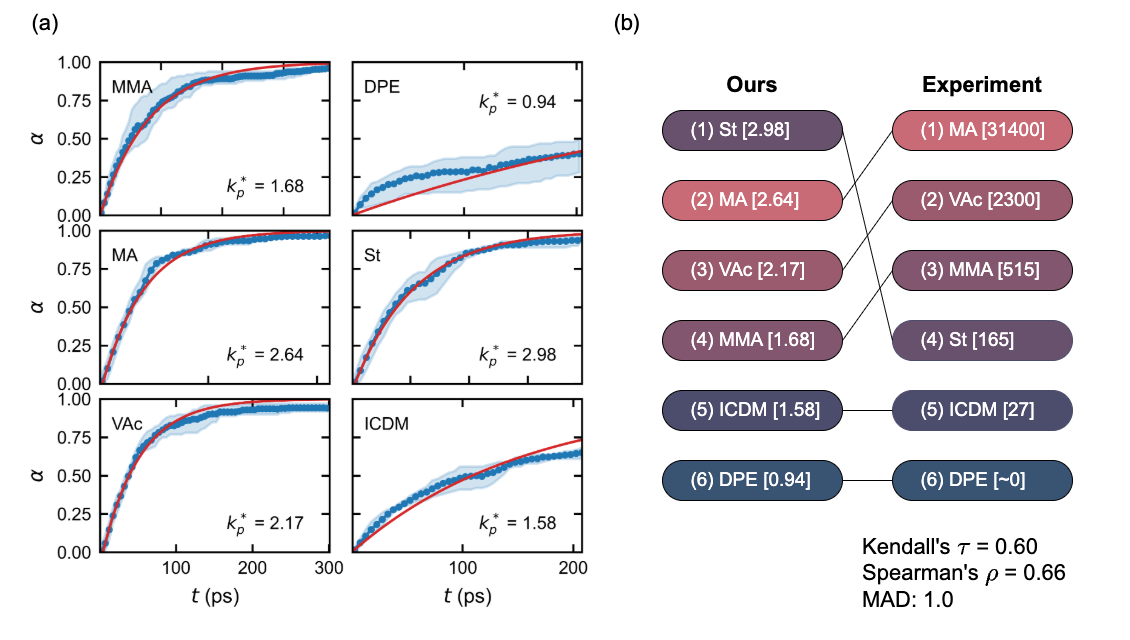}
    \caption{(a) Time--conversion profiles (blue) and exponential fits (red) for representative vinyl monomers, with extracted apparent propagation rate constants $k_p$. 
(b) Ranking of the apparent $k_p$ values compared with experimental data. Rank correlations indicate Kendall’s $\tau=0.60$, Spearman’s $\rho=0.66$, and mean absolute deviation = 1.0.}
    \label{fig:vinyl}
\end{figure*}

Figure~\ref{fig:vinyl} presents the apparent propagation rate constants ($k_p^*$) for various vinyl monomers. 
The apparent rate constants were obtained by fitting the monomer conversion profiles to a single-exponential function,
\begin{equation}
\alpha(t) = 1 - \exp(-k_p^\ast t),
\end{equation}
where $\alpha(t)$ is the monomer conversion at time $t$, and $k_p^\ast$ represents the effective propagation rate under the simulated conditions. 
Representative fits are shown in Figure~\ref{fig:vinyl} (a), together with the extracted $k_p$ values for each monomer. 
These fitted values were then ranked and compared against experimental propagation rate constants (Figure~\ref{fig:vinyl} (b)). 
Rank correlations between the simulated and experimentally reported propagation rates yield a Spearman correlation coefficient ($\rho$) of 0.66, Kendall rank correlation ($\tau$) of 0.60, and mean absolute deviation of 1.0. While the agreement is not perfect, the simulations provide a reasonable basis for ranking relative monomer reactivities.

To probe the origin of discrepancies in ranking, explicit transition-state searches were performed using the PFP to estimate activation energies for representative propagation reactions (Table~S4). 
Although experimental activation energies place styrene in the middle of the monomer series, the barriers computed using PFP rank styrene as the lowest among the studied monomers. 
This inconsistency in energetic ordering parallels the deviation observed in the simulated propagation rate constants, indicating that the adopted potential systematically underestimates the propagation barrier of styrene. 
In contrast, bulk densities are accurately reproduced (Table~S4), suggesting that the discrepancy originates from the description of reaction barriers rather than from the condensed-phase structural properties. 
Improving the barrier fidelity of the PFP is therefore expected to directly enhance the reliability of kinetic rankings.

Nonetheless, rank inversions are observed for certain monomers, highlighting areas where the proposed model struggles to capture subtle differences in propagation kinetics. 
The range of propagation rate constants examined corresponds to activation energy variations of only a few kcal/mol.
Accurately resolving such fine energetic differences will likely require higher-fidelity uMLIP trained on more comprehensive and accurate reference data, or advanced acceleration schemes capable of capturing the delicate energetics of reactive events.
Notably, because the applied bias alters the effective timescale of the dynamics, the proposed approach is not intended to reproduce absolute propagation rate constants. 
Rather, its primary utility lies in capturing relative trends and reactivity rankings across monomers without requiring prior knowledge of activation energies. 
Incorporating higher-fidelity potentials or refined acceleration strategies is expected to further improve quantitative agreement and expand the applicability of this framework to complex polymerization systems.

\subsection{Step-Growth Polycondensation of Nylon-6,6}
\begin{figure*}
    \centering
    \includegraphics[width=0.9\linewidth]{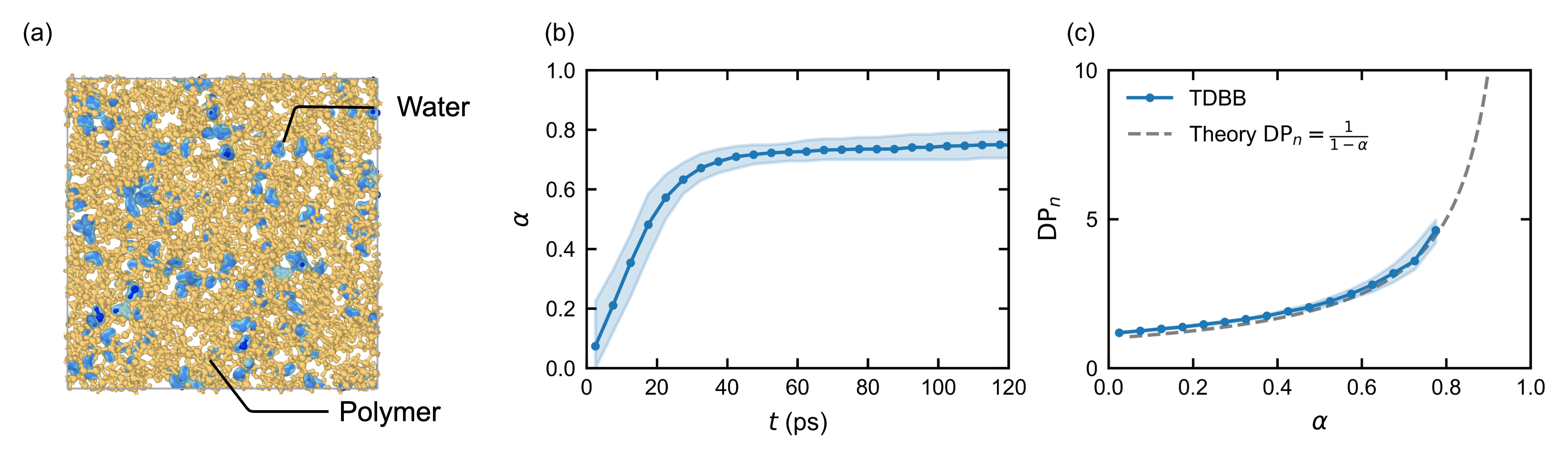}
    \caption{Structural and kinetic characteristics of nylon-6,6 step-growth polymerization under melt conditions without water removal. 
    (a) Representative snapshot after polymerization, showing polymer chains (tan) and generated water molecules (blue). 
    (b) Time evolution of monomer conversion ($\alpha$), which levels off below unity owing to equilibrium limitation. 
    (c) Number-average degree of polymerization (DP$_n$) as a function of conversion, compared with the theoretical Carothers relation. 
    The simulations reproduce the characteristic nonlinear DP$_n$--conversion relationship and incomplete conversion observed experimentally in condensation polymerizations without continuous removal of condensate.}
    \label{fig:nylon}
\end{figure*}

To further evaluate the capability of the proposed framework in modeling step-growth polymerization, we examined the polymerization of nylon-6,6, a prototypical condensation polymer. The results (Figure~\ref{fig:nylon}) show the time evolution of monomer conversion and the relationship between conversion and DP$_n$. Here, DP$_n$ denotes the number-average degree of polymerization, defined as the average number of monomeric units per polymer chain. DP$_n$ is related to the number-average molar mass $M_\mathrm{n}$ by $\mathrm{DP}_n = M_\mathrm{n}/M_0$, where $M_0$ is the molar mass of the repeating unit. Consistent with classical theories of step-growth polymerization, DP$_n$ increases monotonically with conversion, particularly in the later stages dominated by oligomer–oligomer coupling. This trend, a hallmark feature of step-growth kinetics, is reproduced with high fidelity across all runs, underscoring the robustness of the methodology. The quantitative agreement between simulation and theoretical expectations indicates that the uMLIP–accelerated framework accurately captures the essential thermodynamic and kinetic characteristics of condensation polymerizations.

Our results also align with experimental observations of nylon polycondensation. In melt polymerization without continuous water removal, conversion typically levels off at 60–80\%.\cite{odian2004principles} Our simulations reproduce this incomplete conversion and limited chain length, in contrast to idealized models that assume perfect removal of the condensate. This correspondence highlights the ability of the proposed framework to capture the equilibrium-limited nature of step-growth polymerizations without requiring predefined activation barriers or frequency factors for each reaction pathway, demonstrating a clear advantage over heuristic methods.

\subsection{Epoxy Curing at the CuO Interface}

Reliable adhesion between epoxy resins and metallic wiring layers is essential for improving the performance and reliability of semiconductor devices. 
Surface treatments such as oxygen plasma are commonly applied to enhance adhesion, and spectroscopic studies have revealed the presence of Cu--O--C linkages at treated interfaces.\cite{duguet2019dft} 
These observations suggest that covalent bond formation across the epoxy--copper interface is a key factor underlying the improved interfacial strength. 
However, molecular-level understanding of these interfacial curing reactions remains challenging to achieve owing to their chemical complexity and large system sizes.

This section describes the assessment of the ability of the proposed accelerated simulation framework to capture interfacial curing chemistry. 
The considered case is particularly demanding: No publicly available ReaxFF parametrization simultaneously covers Cu surfaces and epoxy--amine chemistry, and full ab initio dynamics of interfacial systems of this scale are computationally infeasible. 
Consequently, conventional approaches are not directly applicable, and uMLIPs provide the only practical route to modeling these interfaces. 
Even so, straightforward uMLIP MD suffers from the rare-event bottleneck, failing to capture curing reactions within accessible time windows. 

\begin{figure*}[htbp]
    \centering
    \includegraphics[width=1.0\linewidth]{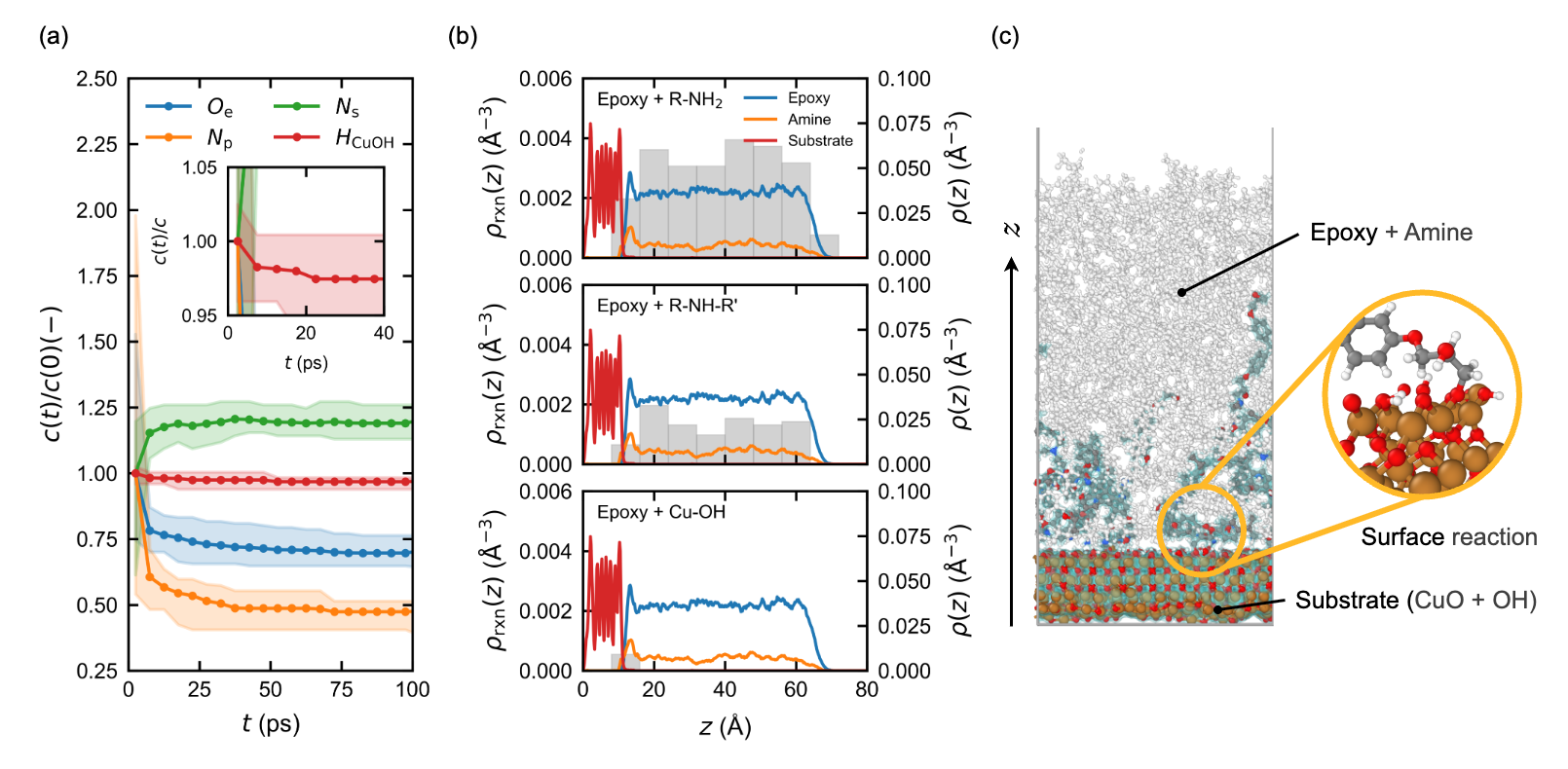}
    \caption{
Epoxy curing at the hydroxylated CuO interface.
(a) Time evolution of relative concentrations $c(t)/c(0)$ of epoxy oxygen ($O_\mathrm{e}$), 
primary amine nitrogen ($N_\mathrm{p}$), secondary amine nitrogen ($N_\mathrm{s}$), 
and surface hydroxyl hydrogen ($H_\mathrm{CuOH}$).
(b) Depth-resolved reaction density $\rho_{\mathrm{rxn}}(z)$ for epoxy--primary amine (top), 
epoxy--secondary amine (middle), and epoxy--surface reactions (bottom), 
together with atomic number density profiles $\rho(z)$ of epoxy (blue), amine (orange), 
and CuO substrate (red).
(c) Representative snapshot of the epoxy/CuO interface, with magnified view showing 
covalent bond formation between epoxy and surface hydroxyl groups.
}
    \label{fig:epoxycopper}
\end{figure*}

To analyze the curing reaction, 
Figure~\ref{fig:epoxycopper}(a) shows the relative concentration change 
$c(t)/c(0)$ of key atomic species as a function of simulation time:
\[
c(t)/c(0) = \frac{N_\mathrm{species}(t)}{N_\mathrm{species}(0)}.
\]
Here, $N_\mathrm{species}(t)$ is the number of target atoms at time $t$.
Both epoxy oxygen ($O_\mathrm{e}$) and primary amine nitrogen ($N_\mathrm{p}$) 
decrease over time, while secondary amine nitrogen ($N_\mathrm{s}$) 
increases. This indicates that epoxy ring-opening by primary amines is the 
dominant reaction in the early stage. The increase in $N_\mathrm{s}$ confirms 
the formation of secondary amines, consistent with the expected 
initial step of epoxy curing. Experimental studies of diglycidyl ether of bisphenol A/diethylenetriamine systems 
using near-infrared spectroscopy reveal that primary amines are nearly consumed 
before secondary amine conversion begins. Our simulations capture this 
early-stage behavior. The number of hydroxyl hydrogens on the CuO surface 
($H_\mathrm{CuOH}$) decreases slightly, suggesting limited reaction at the 
interface.

To further characterize the spatial distribution of reaction events, we evaluated the depth-resolved reaction density $\rho_{\mathrm{rxn}}(z)$:
\[
\rho_{\mathrm{rxn}}(z) =
\frac{N_{\mathrm{rxn}}(z)}{A\,\Delta z\,N_{\mathrm{frames}}},
\]
where $N_{\mathrm{rxn}}(z)$ is the number of reactions observed in bin $z$, 
$A$ is the cross-sectional area, and $\Delta z$ is the bin width, set to 8~\AA. 
Figure~\ref{fig:epoxycopper}(b) shows $\rho_{\mathrm{rxn}}(z)$ for epoxy--primary 
amine, epoxy--secondary amine, and epoxy--surface reactions, together with 
the atomic number density profiles $\rho(z)$ of epoxy, amine, and substrate. 
Interestingly, the reaction density near the interface is lower than in 
the bulk, even though the local atomic densities of both epoxy and amine 
are higher near the interface. This finding is consistent with prior 
studies reporting densification near solid substrates, where restricted 
segmental mobility leads to reduced reaction rates despite high local 
concentrations. 
In other words, our simulations confirm that interfacial reactions are 
suppressed in the near-surface region, in agreement with previous 
experimental and theoretical reports of reduced reactivity at polymer/solid 
interfaces. \cite{yamamoto2020molecular}

A representative snapshot is shown in Figure~\ref{fig:epoxycopper}(c).
Direct visualization confirms that epoxy ring-opening reactions occur 
at the CuO--OH surface, leading to covalent bond formation between the 
epoxy network and substrate. Even though the number of such 
interfacial bonds is small, their formation connects the polymer network 
to the substrate and is expected to significantly enhance adhesion, 
consistent with experimental observations.

\section{Conclusion}

The proposed framework successfully captures relative reactivity trends and 
qualitative mechanistic features. However, several limitations remain to be addressed. 
First, applying a bias potential disrupts the correspondence between 
simulation and physical time, preventing direct reproduction of absolute rate constants. 
Second, although propagation events are systematically accelerated, the current 
implementation excludes other relevant pathways, such as chain transfer and termination, 
which are essential for describing molecular-weight distributions in radical polymerization. 
Third, the accuracy of predicted barriers and kinetics is ultimately bounded by the 
fidelity of the underlying uMLIP, which, in our case, does not fully 
resolve subtle energetic differences across monomers.

Thus, the following directions are recommended for future work: 
(i) development of higher-fidelity uMLIPs trained on more diverse and accurate reference data, 
including reaction barriers and transition-state information, to improve quantitative 
reliability; 
(ii) extension of the reaction catalog to incorporate side reactions and termination events, 
enabling a more comprehensive description of experimentally observed polymerization kinetics; and 
(iii) exploration of hybrid approaches that combine acceleration with approximate time 
rescaling or kinetic reconstruction, thereby bridging the gap between qualitative 
trend capture and quantitative rate prediction. 
In addition, coupling the present scheme with enhanced conformational sampling 
(e.g., replica exchange or basin-hopping strategies) may further improve the robustness of 
predicted mechanistic pathways.

Overall, the TDBB approach, combined with a transferable uMLIP, provides a practical 
and versatile platform for simulating polymerization and curing processes. 
Its demonstrated ability to reproduce key features across diverse systems without 
system-specific parametrization highlights its promise as a foundation for developing more
quantitatively predictive and chemically comprehensive methodologies.

\begin{suppinfo}
Additional figures, reaction pathway analyses, and simulation details
are provided in the Supporting Information (PDF) available at the ACS Publications website.
\end{suppinfo}

\begin{acknowledgement}
We express our sincere gratitude to So Takamoto of Preferred Networks, Inc., and Ryosuke Kuwabara of MITSUI \& CO., LTD., and Sannohe Kunio of IBLC Co., Ltd. as well as to all their colleagues, for their valuable comments, insightful discussions, and continuous support.
\end{acknowledgement}
\bibliography{references}

\providecommand{\latin}[1]{#1}
\makeatletter
\providecommand{\doi}
  {\begingroup\let\do\@makeother\dospecials
  \catcode`\{=1 \catcode`\}=2 \doi@aux}
\providecommand{\doi@aux}[1]{\endgroup\texttt{#1}}
\makeatother
\providecommand*\mcitethebibliography{\thebibliography}
\csname @ifundefined\endcsname{endmcitethebibliography}  {\let\endmcitethebibliography\endthebibliography}{}
\begin{mcitethebibliography}{50}
\providecommand*\natexlab[1]{#1}
\providecommand*\mciteSetBstSublistMode[1]{}
\providecommand*\mciteSetBstMaxWidthForm[2]{}
\providecommand*\mciteBstWouldAddEndPuncttrue
  {\def\EndOfBibitem{\unskip.}}
\providecommand*\mciteBstWouldAddEndPunctfalse
  {\let\EndOfBibitem\relax}
\providecommand*\mciteSetBstMidEndSepPunct[3]{}
\providecommand*\mciteSetBstSublistLabelBeginEnd[3]{}
\providecommand*\EndOfBibitem{}
\mciteSetBstSublistMode{f}
\mciteSetBstMaxWidthForm{subitem}{(\alph{mcitesubitemcount})}
\mciteSetBstSublistLabelBeginEnd
  {\mcitemaxwidthsubitemform\space}
  {\relax}
  {\relax}

\bibitem[McCrum \latin{et~al.}(1997)McCrum, Buckley, and Bucknall]{McCrum1997-ge}
McCrum,~N.~G.; Buckley,~C.~P.; Bucknall,~C.~B. \emph{Principles of polymer engineering}; Oxford University Press, 1997\relax
\mciteBstWouldAddEndPuncttrue
\mciteSetBstMidEndSepPunct{\mcitedefaultmidpunct}
{\mcitedefaultendpunct}{\mcitedefaultseppunct}\relax
\EndOfBibitem
\bibitem[Ebewele(2000)]{Ebewele2000-fl}
Ebewele,~R.~O. \emph{Polymer Science and Technology}; CRC Press, 2000\relax
\mciteBstWouldAddEndPuncttrue
\mciteSetBstMidEndSepPunct{\mcitedefaultmidpunct}
{\mcitedefaultendpunct}{\mcitedefaultseppunct}\relax
\EndOfBibitem
\bibitem[Young and Lovell(2011)Young, and Lovell]{Young2011-oz}
Young,~R.~J.; Lovell,~P.~A. \emph{Introduction to Polymers}; CRC Press, 2011\relax
\mciteBstWouldAddEndPuncttrue
\mciteSetBstMidEndSepPunct{\mcitedefaultmidpunct}
{\mcitedefaultendpunct}{\mcitedefaultseppunct}\relax
\EndOfBibitem
\bibitem[Wypych(2022)]{Wypych2022-lf}
Wypych,~G. \emph{Handbook of polymers}; Elsevier, 2022\relax
\mciteBstWouldAddEndPuncttrue
\mciteSetBstMidEndSepPunct{\mcitedefaultmidpunct}
{\mcitedefaultendpunct}{\mcitedefaultseppunct}\relax
\EndOfBibitem
\bibitem[Rao \latin{et~al.}(2022)Rao, Takayanagi, and Nagaoka]{Rao2022-jj}
Rao,~Z.; Takayanagi,~M.; Nagaoka,~M. Verification for Temperature Dependence of Tacticity in Polystyrene Radical Polymerization with the Combination of Reaction Pathway Analysis and Red Moon Methodology. \emph{J. Phys. Chem. B} \textbf{2022}, \emph{126}, 5343--5350\relax
\mciteBstWouldAddEndPuncttrue
\mciteSetBstMidEndSepPunct{\mcitedefaultmidpunct}
{\mcitedefaultendpunct}{\mcitedefaultseppunct}\relax
\EndOfBibitem
\bibitem[Bezik \latin{et~al.}(2023)Bezik, Redline, Foster, and Frischknecht]{Bezik2023-ts}
Bezik,~C.~T.; Redline,~E.~M.; Foster,~J.~C.; Frischknecht,~A.~L. Simulations of glass transition and mechanical behavior of off-stoichiometric crosslinked polymers. \emph{Macromolecules} \textbf{2023}, \emph{56}, 5268--5277\relax
\mciteBstWouldAddEndPuncttrue
\mciteSetBstMidEndSepPunct{\mcitedefaultmidpunct}
{\mcitedefaultendpunct}{\mcitedefaultseppunct}\relax
\EndOfBibitem
\bibitem[Pramanik \latin{et~al.}(2014)Pramanik, Fowler, and Rawlins]{Pramanik2014-uy}
Pramanik,~M.; Fowler,~E.~W.; Rawlins,~J.~W. Cure kinetics of several epoxy–amine systems at ambient and high temperatures. \emph{J Coat Technol Res} \textbf{2014}, \emph{11}, 143--157\relax
\mciteBstWouldAddEndPuncttrue
\mciteSetBstMidEndSepPunct{\mcitedefaultmidpunct}
{\mcitedefaultendpunct}{\mcitedefaultseppunct}\relax
\EndOfBibitem
\bibitem[Beuermann and Buback(2002)Beuermann, and Buback]{Beuermann2002-ps}
Beuermann,~S.; Buback,~M. Rate coefficients of free-radical polymerization deduced from pulsed laser experiments. \emph{Prog. Polym. Sci.} \textbf{2002}, \emph{27}, 191--254\relax
\mciteBstWouldAddEndPuncttrue
\mciteSetBstMidEndSepPunct{\mcitedefaultmidpunct}
{\mcitedefaultendpunct}{\mcitedefaultseppunct}\relax
\EndOfBibitem
\bibitem[Gartner and Jayaraman(2019)Gartner, and Jayaraman]{Gartner2019-lq}
Gartner,~T.~E.,~III; Jayaraman,~A. Modeling and simulations of polymers: A roadmap. \emph{Macromolecules} \textbf{2019}, \emph{52}, 755--786\relax
\mciteBstWouldAddEndPuncttrue
\mciteSetBstMidEndSepPunct{\mcitedefaultmidpunct}
{\mcitedefaultendpunct}{\mcitedefaultseppunct}\relax
\EndOfBibitem
\bibitem[Joshi and Deshmukh(2021)Joshi, and Deshmukh]{Joshi2021-kh}
Joshi,~S.~Y.; Deshmukh,~S.~A. A review of advancements in coarse-grained molecular dynamics simulations. \emph{Mol. Simul.} \textbf{2021}, \emph{47}, 786--803\relax
\mciteBstWouldAddEndPuncttrue
\mciteSetBstMidEndSepPunct{\mcitedefaultmidpunct}
{\mcitedefaultendpunct}{\mcitedefaultseppunct}\relax
\EndOfBibitem
\bibitem[Krishna \latin{et~al.}(2021)Krishna, Sreedhar, and Patel]{Krishna2021-ku}
Krishna,~S.; Sreedhar,~I.; Patel,~C.~M. Molecular dynamics simulation of polyamide-based materials – A review. \emph{Comput. Mater. Sci.} \textbf{2021}, \emph{200}, 110853\relax
\mciteBstWouldAddEndPuncttrue
\mciteSetBstMidEndSepPunct{\mcitedefaultmidpunct}
{\mcitedefaultendpunct}{\mcitedefaultseppunct}\relax
\EndOfBibitem
\bibitem[Liu \latin{et~al.}(2021)Liu, Xu, Wang, Yang, Duan, Ma, Lin, and Liu]{Liu2021-og}
Liu,~Z.; Xu,~Z.; Wang,~D.; Yang,~Y.; Duan,~Y.; Ma,~L.; Lin,~T.; Liu,~H. A review on molecularly imprinted polymers preparation by computational simulation-aided methods. \emph{Polymers} \textbf{2021}, \emph{13}\relax
\mciteBstWouldAddEndPuncttrue
\mciteSetBstMidEndSepPunct{\mcitedefaultmidpunct}
{\mcitedefaultendpunct}{\mcitedefaultseppunct}\relax
\EndOfBibitem
\bibitem[Arash \latin{et~al.}(2017)Arash, Thijsse, Pecenko, and Simone]{Arash2017-oi}
Arash,~B.; Thijsse,~B.~J.; Pecenko,~A.; Simone,~A. Effect of water content on the thermal degradation of amorphous polyamide 6,6: A collective variable-driven hyperdynamics study. \emph{Polym. Degrad. Stab.} \textbf{2017}, \emph{146}, 260--266\relax
\mciteBstWouldAddEndPuncttrue
\mciteSetBstMidEndSepPunct{\mcitedefaultmidpunct}
{\mcitedefaultendpunct}{\mcitedefaultseppunct}\relax
\EndOfBibitem
\bibitem[Mori and Matubayasi(2018)Mori, and Matubayasi]{Mori2018-be}
Mori,~H.; Matubayasi,~N. Resin filling into nano-sized pore on metal surface analyzed by all-atom molecular dynamics simulation over a variety of resin and pore sizes. \emph{Polymer} \textbf{2018}, \emph{150}, 360--370\relax
\mciteBstWouldAddEndPuncttrue
\mciteSetBstMidEndSepPunct{\mcitedefaultmidpunct}
{\mcitedefaultendpunct}{\mcitedefaultseppunct}\relax
\EndOfBibitem
\bibitem[Odegard \latin{et~al.}(2021)Odegard, Patil, Deshpande, Kanhaiya, Winetrout, Heinz, Shah, and Maiaru]{Odegard2021-nc}
Odegard,~G.~M.; Patil,~S.~U.; Deshpande,~P.~P.; Kanhaiya,~K.; Winetrout,~J.~J.; Heinz,~H.; Shah,~S.~P.; Maiaru,~M. Molecular dynamics modeling of epoxy resins using the reactive interface force field. \emph{Macromolecules} \textbf{2021}, \emph{54}, 9815--9824\relax
\mciteBstWouldAddEndPuncttrue
\mciteSetBstMidEndSepPunct{\mcitedefaultmidpunct}
{\mcitedefaultendpunct}{\mcitedefaultseppunct}\relax
\EndOfBibitem
\bibitem[Yamaguchi \latin{et~al.}(2022)Yamaguchi, Kawaguchi, Miyata, Miyazaki, Aoki, Yamamoto, and Tanaka]{Yamaguchi2022-je}
Yamaguchi,~K.; Kawaguchi,~D.; Miyata,~N.; Miyazaki,~T.; Aoki,~H.; Yamamoto,~S.; Tanaka,~K. Kinetics of the interfacial curing reaction for an epoxy-amine mixture. \emph{Phys. Chem. Chem. Phys.} \textbf{2022}, \emph{24}, 21578--21582\relax
\mciteBstWouldAddEndPuncttrue
\mciteSetBstMidEndSepPunct{\mcitedefaultmidpunct}
{\mcitedefaultendpunct}{\mcitedefaultseppunct}\relax
\EndOfBibitem
\bibitem[Okabe \latin{et~al.}(2013)Okabe, Takehara, Inose, Hirano, Nishikawa, and Uehara]{Okabe2013-lk}
Okabe,~T.; Takehara,~T.; Inose,~K.; Hirano,~N.; Nishikawa,~M.; Uehara,~T. Curing reaction of epoxy resin composed of mixed base resin and curing agent: Experiments and molecular simulation. \emph{Polymer} \textbf{2013}, \emph{54}, 4660--4668\relax
\mciteBstWouldAddEndPuncttrue
\mciteSetBstMidEndSepPunct{\mcitedefaultmidpunct}
{\mcitedefaultendpunct}{\mcitedefaultseppunct}\relax
\EndOfBibitem
\bibitem[Nagaoka \latin{et~al.}(2013)Nagaoka, Suzuki, Okamoto, and Takenaka]{nagaoka2013hybrid}
Nagaoka,~M.; Suzuki,~Y.; Okamoto,~T.; Takenaka,~N. A hybrid MC/MD reaction method with rare event-driving mechanism: Atomistic realization of 2-chlorobutane racemization process in DMF solution. \emph{Chemical Physics Letters} \textbf{2013}, \emph{583}, 80--86\relax
\mciteBstWouldAddEndPuncttrue
\mciteSetBstMidEndSepPunct{\mcitedefaultmidpunct}
{\mcitedefaultendpunct}{\mcitedefaultseppunct}\relax
\EndOfBibitem
\bibitem[Suzuki and Nagaoka(2017)Suzuki, and Nagaoka]{suzuki2017transformation}
Suzuki,~Y.; Nagaoka,~M. A transformation theory of stochastic evolution in Red Moon methodology to time evolution of chemical reaction process in the full atomistic system. \emph{The Journal of Chemical Physics} \textbf{2017}, \emph{146}\relax
\mciteBstWouldAddEndPuncttrue
\mciteSetBstMidEndSepPunct{\mcitedefaultmidpunct}
{\mcitedefaultendpunct}{\mcitedefaultseppunct}\relax
\EndOfBibitem
\bibitem[Gissinger \latin{et~al.}(2017)Gissinger, Jensen, and Wise]{Gissinger2017-wh}
Gissinger,~J.~R.; Jensen,~B.~D.; Wise,~K.~E. Chemical reactions in classical molecular dynamics. \emph{Polymer} \textbf{2017}, \emph{128}, 211--217\relax
\mciteBstWouldAddEndPuncttrue
\mciteSetBstMidEndSepPunct{\mcitedefaultmidpunct}
{\mcitedefaultendpunct}{\mcitedefaultseppunct}\relax
\EndOfBibitem
\bibitem[Gissinger \latin{et~al.}(2024)Gissinger, Jensen, and Wise]{Gissinger2024-nx}
Gissinger,~J.~R.; Jensen,~B.~D.; Wise,~K.~E. Molecular modeling of reactive systems with {REACTER}. \emph{Comput. Phys. Commun.} \textbf{2024}, \emph{304}, 109287\relax
\mciteBstWouldAddEndPuncttrue
\mciteSetBstMidEndSepPunct{\mcitedefaultmidpunct}
{\mcitedefaultendpunct}{\mcitedefaultseppunct}\relax
\EndOfBibitem
\bibitem[Xi \latin{et~al.}(2024)Xi, Fukuzawa, Kikugawa, Zhao, Kawagoe, Okabe, Kishi, and Kishimoto]{Xi2024-hl}
Xi,~Y.; Fukuzawa,~H.; Kikugawa,~G.; Zhao,~Y.; Kawagoe,~Y.; Okabe,~T.; Kishi,~H.; Kishimoto,~N. Enhancing epoxy resin curing: Investigating the catalytic role of water as a trace impurity in dense crosslinked network formation using an advanced cat-{GRRM}/{MC}/{MD} {Method1}. \emph{Polymer (Guildf.)} \textbf{2024}, \emph{313}, 127675\relax
\mciteBstWouldAddEndPuncttrue
\mciteSetBstMidEndSepPunct{\mcitedefaultmidpunct}
{\mcitedefaultendpunct}{\mcitedefaultseppunct}\relax
\EndOfBibitem
\bibitem[Xi \latin{et~al.}(2024)Xi, Fukuzawa, Fukunaga, Kikugawa, Zhao, Kawagoe, Okabe, and Kishimoto]{Xi2024-lq}
Xi,~Y.; Fukuzawa,~H.; Fukunaga,~S.; Kikugawa,~G.; Zhao,~Y.; Kawagoe,~Y.; Okabe,~T.; Kishimoto,~N. Development of cat-{GRRM}/{MC}/{MD} method for the simulation of cross-linked network structure formation with molecular autocatalysis. \emph{Molecular Catalysis} \textbf{2024}, \emph{552}, 113680\relax
\mciteBstWouldAddEndPuncttrue
\mciteSetBstMidEndSepPunct{\mcitedefaultmidpunct}
{\mcitedefaultendpunct}{\mcitedefaultseppunct}\relax
\EndOfBibitem
\bibitem[van Duin \latin{et~al.}(2001)van Duin, Dasgupta, Lorant, and Goddard]{Van_Duin2001-ag}
van Duin,~A. C.~T.; Dasgupta,~S.; Lorant,~F.; Goddard,~W.~A. {ReaxFF}: A Reactive Force Field for Hydrocarbons. \emph{J. Phys. Chem. A} \textbf{2001}, \emph{105}, 9396--9409\relax
\mciteBstWouldAddEndPuncttrue
\mciteSetBstMidEndSepPunct{\mcitedefaultmidpunct}
{\mcitedefaultendpunct}{\mcitedefaultseppunct}\relax
\EndOfBibitem
\bibitem[Brandrup \latin{et~al.}(1999)Brandrup, Immergut, Grulke, Abe, and Bloch]{brandrup1999polymer}
Brandrup,~J.; Immergut,~E.~H.; Grulke,~E.~A.; Abe,~A.; Bloch,~D.~R. \emph{Polymer handbook}; Wiley New York, 1999; Vol.~89\relax
\mciteBstWouldAddEndPuncttrue
\mciteSetBstMidEndSepPunct{\mcitedefaultmidpunct}
{\mcitedefaultendpunct}{\mcitedefaultseppunct}\relax
\EndOfBibitem
\bibitem[Vashisth \latin{et~al.}(2018)Vashisth, Ashraf, Zhang, Bakis, and van Duin]{Vashisth2018-kb}
Vashisth,~A.; Ashraf,~C.; Zhang,~W.; Bakis,~C.~E.; van Duin,~A. C.~T. Accelerated {ReaxFF} Simulations for Describing the Reactive Cross-Linking of Polymers. \emph{J. Phys. Chem. A} \textbf{2018}, \emph{122}, 6633--6642\relax
\mciteBstWouldAddEndPuncttrue
\mciteSetBstMidEndSepPunct{\mcitedefaultmidpunct}
{\mcitedefaultendpunct}{\mcitedefaultseppunct}\relax
\EndOfBibitem
\bibitem[Dasgupta \latin{et~al.}(2020)Dasgupta, Yilmaz, and van Duin]{Dasgupta2020-ll}
Dasgupta,~N.; Yilmaz,~D.~E.; van Duin,~A. Simulations of the biodegradation of citrate-based polymers for artificial scaffolds using accelerated reactive molecular dynamics. \emph{J. Phys. Chem. B} \textbf{2020}, \emph{124}, 5311--5322\relax
\mciteBstWouldAddEndPuncttrue
\mciteSetBstMidEndSepPunct{\mcitedefaultmidpunct}
{\mcitedefaultendpunct}{\mcitedefaultseppunct}\relax
\EndOfBibitem
\bibitem[Laio and Parrinello(2002)Laio, and Parrinello]{Laio2002-kx}
Laio,~A.; Parrinello,~M. Escaping free-energy minima. \emph{Proc. Natl. Acad. Sci. U. S. A.} \textbf{2002}, \emph{99}, 12562--12566\relax
\mciteBstWouldAddEndPuncttrue
\mciteSetBstMidEndSepPunct{\mcitedefaultmidpunct}
{\mcitedefaultendpunct}{\mcitedefaultseppunct}\relax
\EndOfBibitem
\bibitem[Invernizzi and Parrinello(2020)Invernizzi, and Parrinello]{Invernizzi2020-dg}
Invernizzi,~M.; Parrinello,~M. Rethinking metadynamics: From bias potentials to probability distributions. \emph{J. Phys. Chem. Lett.} \textbf{2020}, \emph{11}, 2731--2736\relax
\mciteBstWouldAddEndPuncttrue
\mciteSetBstMidEndSepPunct{\mcitedefaultmidpunct}
{\mcitedefaultendpunct}{\mcitedefaultseppunct}\relax
\EndOfBibitem
\bibitem[Käser \latin{et~al.}(2023)Käser, Vazquez-Salazar, Meuwly, and Töpfer]{Kaser2023-la}
Käser,~S.; Vazquez-Salazar,~L.~I.; Meuwly,~M.; Töpfer,~K. Neural network potentials for chemistry: concepts, applications and prospects. \emph{Digit. Discov.} \textbf{2023}, \emph{2}, 28--58\relax
\mciteBstWouldAddEndPuncttrue
\mciteSetBstMidEndSepPunct{\mcitedefaultmidpunct}
{\mcitedefaultendpunct}{\mcitedefaultseppunct}\relax
\EndOfBibitem
\bibitem[Duignan(2024)]{Duignan2024-vm}
Duignan,~T.~T. The potential of neural network potentials. \emph{ACS Phys. Chem. Au} \textbf{2024}, \emph{4}, 232--241\relax
\mciteBstWouldAddEndPuncttrue
\mciteSetBstMidEndSepPunct{\mcitedefaultmidpunct}
{\mcitedefaultendpunct}{\mcitedefaultseppunct}\relax
\EndOfBibitem
\bibitem[Martin-Barrios \latin{et~al.}(2024)Martin-Barrios, Navas-Conyedo, Zhang, Chen, and Gulín-González]{Martin-Barrios2024-qa}
Martin-Barrios,~R.; Navas-Conyedo,~E.; Zhang,~X.; Chen,~Y.; Gulín-González,~J. An overview about neural networks potentials in molecular dynamics simulation. \emph{Int. J. Quantum Chem.} \textbf{2024}, \emph{124}\relax
\mciteBstWouldAddEndPuncttrue
\mciteSetBstMidEndSepPunct{\mcitedefaultmidpunct}
{\mcitedefaultendpunct}{\mcitedefaultseppunct}\relax
\EndOfBibitem
\bibitem[Batatia \latin{et~al.}(2022)Batatia, Kov'acs, Simm, Ortner, and Csányi]{Batatia2022-eh}
Batatia,~I.; Kov'acs,~D.; Simm,~G.; Ortner,~C.; Csányi,~G. {MACE}: Higher order equivariant message passing neural networks for fast and accurate force fields. \emph{Neural Inf Process Syst} \textbf{2022}, \emph{abs/2206.07697}, 11423--11436\relax
\mciteBstWouldAddEndPuncttrue
\mciteSetBstMidEndSepPunct{\mcitedefaultmidpunct}
{\mcitedefaultendpunct}{\mcitedefaultseppunct}\relax
\EndOfBibitem
\bibitem[Batatia \latin{et~al.}(2025)Batatia, Batzner, Kovács, Musaelian, Simm, Drautz, Ortner, Kozinsky, and Csányi]{Batatia2025-zd}
Batatia,~I.; Batzner,~S.; Kovács,~D.~P.; Musaelian,~A.; Simm,~G. N.~C.; Drautz,~R.; Ortner,~C.; Kozinsky,~B.; Csányi,~G. The design space of {E}(3)-equivariant atom-centred interatomic potentials. \emph{Nat. Mach. Intell.} \textbf{2025}, \emph{7}, 56--67\relax
\mciteBstWouldAddEndPuncttrue
\mciteSetBstMidEndSepPunct{\mcitedefaultmidpunct}
{\mcitedefaultendpunct}{\mcitedefaultseppunct}\relax
\EndOfBibitem
\bibitem[Takamoto \latin{et~al.}(2022)Takamoto, Shinagawa, Motoki, Nakago, Li, Kurata, Watanabe, Yayama, Iriguchi, Asano, Onodera, Ishii, Kudo, Ono, Sawada, Ishitani, Ong, Yamaguchi, Kataoka, Hayashi, Charoenphakdee, and Ibuka]{Takamoto2022-mj}
Takamoto,~S. \latin{et~al.}  Towards universal neural network potential for material discovery applicable to arbitrary combination of 45 elements. \emph{Nat. Commun.} \textbf{2022}, \emph{13}, 2991\relax
\mciteBstWouldAddEndPuncttrue
\mciteSetBstMidEndSepPunct{\mcitedefaultmidpunct}
{\mcitedefaultendpunct}{\mcitedefaultseppunct}\relax
\EndOfBibitem
\bibitem[Chen and Ong(2022)Chen, and Ong]{Chen2022-en}
Chen,~C.; Ong,~S.~P. A universal graph deep learning interatomic potential for the periodic table. \emph{Nat. Comput. Sci.} \textbf{2022}, \emph{2}, 718--728\relax
\mciteBstWouldAddEndPuncttrue
\mciteSetBstMidEndSepPunct{\mcitedefaultmidpunct}
{\mcitedefaultendpunct}{\mcitedefaultseppunct}\relax
\EndOfBibitem
\bibitem[Deng \latin{et~al.}(2023)Deng, Zhong, Jun, Riebesell, Han, Bartel, and Ceder]{Deng2023-ms}
Deng,~B.; Zhong,~P.; Jun,~K.; Riebesell,~J.; Han,~K.; Bartel,~C.~J.; Ceder,~G. {CHGNet} as a pretrained universal neural network potential for charge-informed atomistic modelling. \emph{Nature Machine Intelligence} \textbf{2023}, \emph{5}, 1031--1041\relax
\mciteBstWouldAddEndPuncttrue
\mciteSetBstMidEndSepPunct{\mcitedefaultmidpunct}
{\mcitedefaultendpunct}{\mcitedefaultseppunct}\relax
\EndOfBibitem
\bibitem[Tayfuroglu \latin{et~al.}(2022)Tayfuroglu, Kocak, and Zorlu]{Tayfuroglu2022-ei}
Tayfuroglu,~O.; Kocak,~A.; Zorlu,~Y. A neural network potential for the {IRMOF} series and its application for thermal and mechanical behaviors. \emph{Phys. Chem. Chem. Phys.} \textbf{2022}, \relax
\mciteBstWouldAddEndPunctfalse
\mciteSetBstMidEndSepPunct{\mcitedefaultmidpunct}
{}{\mcitedefaultseppunct}\relax
\EndOfBibitem
\bibitem[Hisama \latin{et~al.}(2024)Hisama, Ishikawa, Aspera, and Koyama]{Hisama2024-jc}
Hisama,~K.; Ishikawa,~A.; Aspera,~S.~M.; Koyama,~M. Theoretical catalyst screening of multielement alloy catalysts for ammonia synthesis using machine learning potential and generative artificial intelligence. \emph{J. Phys. Chem. C Nanomater. Interfaces} \textbf{2024}, \emph{128}, 18750--18758\relax
\mciteBstWouldAddEndPuncttrue
\mciteSetBstMidEndSepPunct{\mcitedefaultmidpunct}
{\mcitedefaultendpunct}{\mcitedefaultseppunct}\relax
\EndOfBibitem
\bibitem[Hisama \latin{et~al.}(2024)Hisama, Valadez~Huerta, and Koyama]{Hisama2024-ey}
Hisama,~K.; Valadez~Huerta,~G.; Koyama,~M. Molecular dynamics of liquid-electrode interface by integrating Coulomb interaction into universal neural network potential. \emph{J. Comput. Chem.} \textbf{2024}, \emph{45}, 2805--2811\relax
\mciteBstWouldAddEndPuncttrue
\mciteSetBstMidEndSepPunct{\mcitedefaultmidpunct}
{\mcitedefaultendpunct}{\mcitedefaultseppunct}\relax
\EndOfBibitem
\bibitem[Lin \latin{et~al.}(2025)Lin, Otake, Kajiwara, Hiraide, Nurhuda, Packwood, Kadota, Sakamoto, Kawaguchi, Kubota, Yao, Horike, Sun, and Kitagawa]{Lin2025-sw}
Lin,~Z.; Otake,~K.-I.; Kajiwara,~T.; Hiraide,~S.; Nurhuda,~M.; Packwood,~D.; Kadota,~K.; Sakamoto,~H.; Kawaguchi,~S.; Kubota,~Y.; Yao,~M.-S.; Horike,~S.; Sun,~X.; Kitagawa,~S. Interconnected lamellar {3D} semiconductive {PCP} for rechargeable aqueous zinc battery cathodes. \emph{Small} \textbf{2025}, e2411386\relax
\mciteBstWouldAddEndPuncttrue
\mciteSetBstMidEndSepPunct{\mcitedefaultmidpunct}
{\mcitedefaultendpunct}{\mcitedefaultseppunct}\relax
\EndOfBibitem
\bibitem[Voter(1997)]{Voter1997-tf}
Voter,~A.~F. Hyperdynamics: Accelerated molecular dynamics of infrequent events. \emph{Phys. Rev. Lett.} \textbf{1997}, \emph{78}, 3908--3911\relax
\mciteBstWouldAddEndPuncttrue
\mciteSetBstMidEndSepPunct{\mcitedefaultmidpunct}
{\mcitedefaultendpunct}{\mcitedefaultseppunct}\relax
\EndOfBibitem
\bibitem[Miron and Fichthorn(2003)Miron, and Fichthorn]{Miron2003-ed}
Miron,~R.~A.; Fichthorn,~K.~A. Accelerated molecular dynamics with the bond-boost method. \emph{J. Chem. Phys.} \textbf{2003}, \emph{119}, 6210--6216\relax
\mciteBstWouldAddEndPuncttrue
\mciteSetBstMidEndSepPunct{\mcitedefaultmidpunct}
{\mcitedefaultendpunct}{\mcitedefaultseppunct}\relax
\EndOfBibitem
\bibitem[Bal and Neyts(2015)Bal, and Neyts]{Bal2015-kj}
Bal,~K.~M.; Neyts,~E.~C. Merging metadynamics into hyperdynamics: accelerated molecular simulations reaching time scales from microseconds to seconds. \emph{J. Chem. Theory Comput.} \textbf{2015}, \emph{11}, 4545--4554\relax
\mciteBstWouldAddEndPuncttrue
\mciteSetBstMidEndSepPunct{\mcitedefaultmidpunct}
{\mcitedefaultendpunct}{\mcitedefaultseppunct}\relax
\EndOfBibitem
\bibitem[Vashisth \latin{et~al.}(2018)Vashisth, Ashraf, Zhang, Bakis, and Van~Duin]{vashisth2018accelerated}
Vashisth,~A.; Ashraf,~C.; Zhang,~W.; Bakis,~C.~E.; Van~Duin,~A.~C. Accelerated ReaxFF simulations for describing the reactive cross-linking of polymers. \emph{The Journal of Physical Chemistry A} \textbf{2018}, \emph{122}, 6633--6642\relax
\mciteBstWouldAddEndPuncttrue
\mciteSetBstMidEndSepPunct{\mcitedefaultmidpunct}
{\mcitedefaultendpunct}{\mcitedefaultseppunct}\relax
\EndOfBibitem
\bibitem[Eastman \latin{et~al.}(2023)Eastman, Galvelis, Pel{\'a}ez, Abreu, Farr, Gallicchio, Gorenko, Henry, Hu, Huang, \latin{et~al.} others]{eastman2023openmm}
Eastman,~P.; Galvelis,~R.; Pel{\'a}ez,~R.~P.; Abreu,~C.~R.; Farr,~S.~E.; Gallicchio,~E.; Gorenko,~A.; Henry,~M.~M.; Hu,~F.; Huang,~J.; others OpenMM 8: molecular dynamics simulation with machine learning potentials. \emph{The Journal of Physical Chemistry B} \textbf{2023}, \emph{128}, 109--116\relax
\mciteBstWouldAddEndPuncttrue
\mciteSetBstMidEndSepPunct{\mcitedefaultmidpunct}
{\mcitedefaultendpunct}{\mcitedefaultseppunct}\relax
\EndOfBibitem
\bibitem[Odian(2004)]{odian2004principles}
Odian,~G. \emph{Principles of polymerization}; John Wiley \& Sons, 2004\relax
\mciteBstWouldAddEndPuncttrue
\mciteSetBstMidEndSepPunct{\mcitedefaultmidpunct}
{\mcitedefaultendpunct}{\mcitedefaultseppunct}\relax
\EndOfBibitem
\bibitem[Duguet \latin{et~al.}(2019)Duguet, Gavrielides, Esvan, Mineva, and Lacaze-Dufaure]{duguet2019dft}
Duguet,~T.; Gavrielides,~A.; Esvan,~J.; Mineva,~T.; Lacaze-Dufaure,~C. DFT simulation of XPS reveals Cu/epoxy polymer interfacial bonding. \emph{The Journal of Physical Chemistry C} \textbf{2019}, \emph{123}, 30917--30925\relax
\mciteBstWouldAddEndPuncttrue
\mciteSetBstMidEndSepPunct{\mcitedefaultmidpunct}
{\mcitedefaultendpunct}{\mcitedefaultseppunct}\relax
\EndOfBibitem
\bibitem[Yamamoto \latin{et~al.}(2020)Yamamoto, Kuwahara, Aoki, Shundo, and Tanaka]{yamamoto2020molecular}
Yamamoto,~S.; Kuwahara,~R.; Aoki,~M.; Shundo,~A.; Tanaka,~K. Molecular events for an epoxy--amine system at a copper interface. \emph{ACS Applied Polymer Materials} \textbf{2020}, \emph{2}, 1474--1481\relax
\mciteBstWouldAddEndPuncttrue
\mciteSetBstMidEndSepPunct{\mcitedefaultmidpunct}
{\mcitedefaultendpunct}{\mcitedefaultseppunct}\relax
\EndOfBibitem
\end{mcitethebibliography}
\end{document}


\section{Simulation Details}
\label{sec:sim_detail}
Unless otherwise noted, all simulations were performed using OpenMM~8.1.1 with the PreFerred Potential (PFP, v6.0.0), a universal machine learning interatomic potential (uMLIP), under the conditions described in the main text (time step, bias parameters, and thermostat/barostat settings). The following sections summarize system-specific model construction and equilibration procedures.

\subsection{Classical Molecular Dynamics (MD) Setup}
\label{sec:classicmd}
Prior to initiating reactive acceleration MD with the uMLIP, 
all systems were equilibrated using a classical force field to obtain 
physically reasonable starting configurations.

The OpenFF~2.0.0 force field was used for bonded and nonbonded interactions. 
For chemical environments not covered by the standard parameter set, 
supplementary bonded parameters were generated by fitting to 
potential energy surfaces obtained from PFP calculations. 
Partial charges were assigned based on Bader charge analysis with PFP 
and uniformly scaled by a factor of 0.6. 
This procedure yielded values close to AM1-BCC charges 
and provided sufficient accuracy for the initial equilibration.

Electrostatics were treated using the particle-mesh Ewald method. 
A real-space cutoff of 0.9\,nm was used, with a switching function applied from 0.8\,nm.  
van der Waals interactions were truncated at 0.9\,nm, 
and long-range dispersion corrections were applied to the energy and pressure.

Equilibration simulations were performed in the {\it NPT} ensemble 
at 300\,K and 1\,atm. 
Temperature was controlled using a Langevin thermostat with a coupling constant of 1.0\,ps$^{-1}$, 
and pressure was maintained using a Monte Carlo barostat. 
A timestep of 0.5\,fs was employed, and all bonds involving hydrogen atoms were constrained using the RATTLE algorithm. 
Each system was equilibrated for 300\,ps under these conditions.

The equilibrated structures obtained from the classical MD stage 
were subsequently used as initial configurations for production simulations with PFP. 
All results reported in the main text are based on the PFP-based simulations; 
the classical equilibration was used only to prepare well-relaxed starting states.

\subsection{Radical Polymerization of Vinyl Monomers}

Styrene radical polymerization was simulated under various initiator/monomer/solvent (I/M/S) ratios: 
5/100/0, 5/100/100, 1/20/200, and 10/100/0. 
To compare the reactivity of different monomers, systems were prepared with 200 molecules of a given vinyl monomer 
(methyl acrylate, methyl methacrylate, styrene, vinyl acetate, 
diphenylethylene, or dimethyl itaconate) and 10 molecules of azobisisobutyronitrile (AIBN) as the initiator. 
All simulations were performed under solvent-free conditions unless otherwise specified 
(e.g., styrene/toluene mixtures).

\begin{figure}[htbp]
  \centering
  \includegraphics[width=0.8\linewidth]{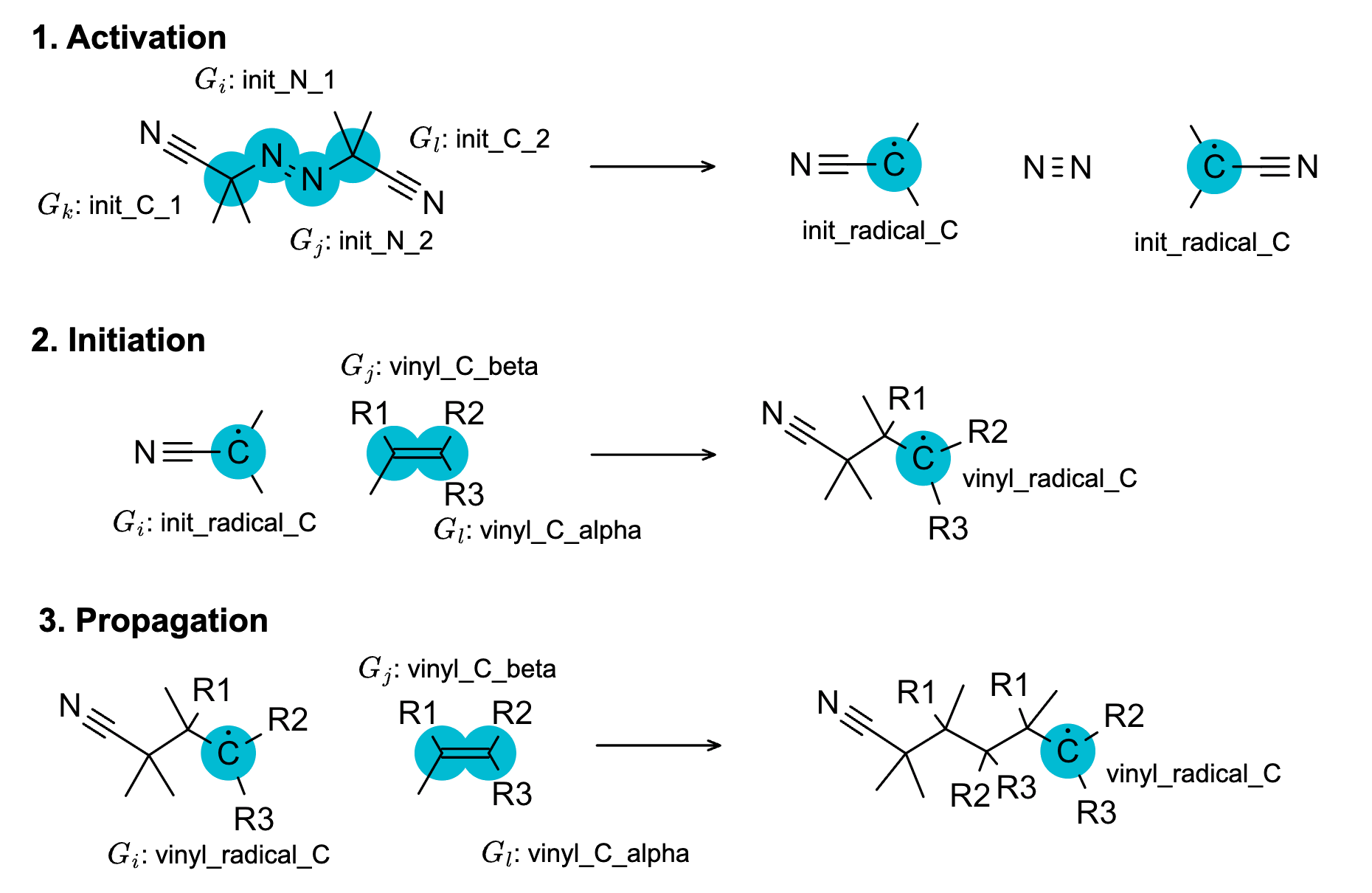}
  \caption{
Reaction patterns considered for vinyl radical polymerization: 
AIBN decomposition, radical initiation, and successive propagation. 
Termination and chain‐transfer reactions were not considered. 
Reactive centers subject to bias are highlighted.
}
  \label{fig:reaction_vinyl}
\end{figure}

Initial configurations were generated at a density of 0.5\,g/mL. 
Systems were first equilibrated using a classical force field (see Section ``Classical Molecular Dynamics (MD) Setup ''), 
followed by a short equilibration with PFP. 
Production simulations using reactive acceleration MD were then carried out at 333\,K and 1\,atm 
in the {\it NPT} ensemble. 
Temperature and pressure were controlled using a Langevin thermostat and Monte Carlo barostat, respectively. 
Only AIBN decomposition and subsequent radical addition to vinyl monomers were enabled 
(Fig.~\ref{fig:reaction_vinyl}), while termination and chain-transfer reactions were excluded.
The corresponding distance criteria and bias assignments for group identification are summarized in 
Table~\ref{tab:reaction_params_vinyl}.

\vspace{2ex}

\begin{table}[h]
\centering
\caption{Search parameters and bias assignment for vinyl radical polymerization. Distances in \AA.}
\label{tab:reaction_params_vinyl}
\begin{tabular}{lcccccccccc}
\toprule
\multirow{4}{*}{Reaction} & \multicolumn{6}{c}{Group identification} & \multicolumn{2}{c}{Applied bias} \\
\cmidrule(lr){2-7} \cmidrule(lr){8-9}
 & \multicolumn{2}{c}{$i$–$j$} & \multicolumn{2}{c}{$i$–$k$} & \multicolumn{2}{c}{$j$–$l$} & $V^{f}$ & $V^{d}$ \\
\cmidrule(lr){2-3} \cmidrule(lr){4-5} \cmidrule(lr){6-7}
 & $r_{\min}$ & $r_{\max}$ & $r_{\min}$ & $r_{\max}$ & $r_{\min}$ & $r_{\max}$ & & \\
\midrule
Activation   & 0.0 & 3.0 & 0.0 & 3.0 & 0.0 & 3.0 &       & $i$–$k$, $j$–$l$ \\
Initiation   & 3.0 & 6.0 & 0.0 & 3.0 & 0.0 & 3.0 & $i$–$j$ &                  \\
Propagation  & 3.0 & 6.0 & 0.0 & 3.0 & 0.0 & 3.0 & $i$–$j$ &                  \\
\bottomrule
\end{tabular}
\end{table}

\subsection{Step-Growth Polycondensation of Nylon-6,6}

Nylon-6,6 polycondensation was simulated starting from an equimolar mixture of 
100 molecules each of hexamethylenediamine and adipic acid. 

\begin{figure}[htbp]
  \centering
  \includegraphics[width=0.8\linewidth]{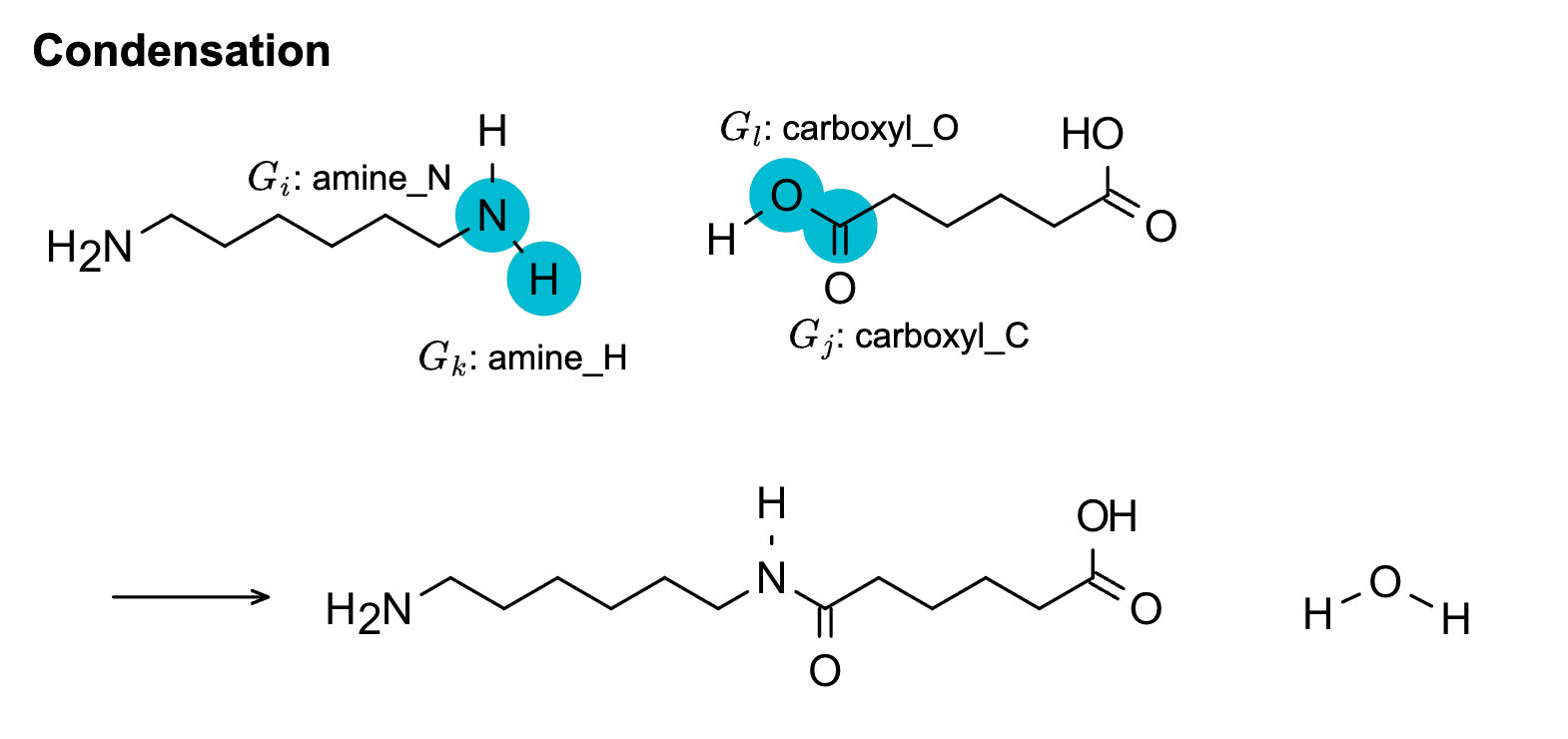}
  \caption{
Reaction pattern for nylon 6,6 step growth polycondensation (amide formation via dehydration). Reactive centers subject to bias are highlighted.
}
  \label{fig:reaction_nylon}
\end{figure}

Initial configurations were generated at a density of 0.5\,g/mL. 
Systems were first equilibrated with a classical force field (see Section ``Classical Molecular Dynamics (MD) Setup''), 
followed by a short equilibration run with PFP. 
Production simulations using reactive acceleration MD were then performed in the {\it NPT} ensemble 
at 300\,K and 1\,atm. 
Temperature and pressure were controlled using a Langevin thermostat and Monte Carlo barostat, respectively. 
Condensation reactions between amine and carboxylic acid groups were selectively accelerated 
to enable amide bond formation, while all other interactions were treated with standard dynamics. 
The reaction patterns considered are presented in Fig.~\ref{fig:reaction_nylon}, 
and the corresponding distance criteria and bias assignments are listed in 
Table~\ref{tab:reaction_params_nylon}.

To ensure statistical reliability, three independent simulations were performed with randomized 
initial configurations, and the reported results represent averages over these runs.

\vspace{2ex}

\begin{table}[h]
\centering
\caption{Search parameters and bias assignment for step-growth polycondensation. Distances in \AA.}
\label{tab:reaction_params_nylon}
\begin{tabular}{lcccccccccc}
\toprule
\multirow{4}{*}{Reaction} & \multicolumn{6}{c}{Group identification} & \multicolumn{2}{c}{Applied bias} \\
\cmidrule(lr){2-7} \cmidrule(lr){8-9}
 & \multicolumn{2}{c}{$i$–$j$} & \multicolumn{2}{c}{$i$–$k$} & \multicolumn{2}{c}{$j$–$l$} & $V^{f}$ & $V^{d}$ \\
\cmidrule(lr){2-3} \cmidrule(lr){4-5} \cmidrule(lr){6-7}
 & $r_{\min}$ & $r_{\max}$ & $r_{\min}$ & $r_{\max}$ & $r_{\min}$ & $r_{\max}$ & & \\
\midrule
Condensation   & 3.0 & 6.0 & 0.0 & 3.0 & 0.0 & 3.0 &  $i$–$j$, $k$–$l$     & $i$–$k$, $j$–$l$ \\
\bottomrule
\end{tabular}
\end{table}

\vspace{2ex}

\subsection{Epoxy Curing at the CuO Interface}

Interface models were constructed by combining a hydroxylated CuO slab with a liquid mixture 
of bisphenol~A diglycidyl ether (DGEBA) and diethylenetriamine (DETA). 
The CuO substrate was prepared by cleaving the (001) surface of bulk CuO, 
constructing an $(8 \times 8 \times 6)$ supercell, and adding a 20~\AA\ vacuum layer along the $z$ direction. 
Fifty hydrogen atoms were randomly placed on the exposed $z$ surface to generate hydroxyl groups. 
The structure was geometry-optimized with PFP, followed by equilibration at 300\,K and 1\,atm for 5\,ps. 

The epoxy/amine mixture consisted of 100 molecules of DGEBA and 50 molecules of DETA, 
packed at an initial density of 0.2\,g/cm$^3$ in a rectangular box with $x$ and $y$ dimensions 
matching those of the slab. 
To efficiently prepare a liquid-like structure, Lennard--Jones wall potentials were applied 
at the top and bottom along the $z$ axis. 
Classical MD (see Section ``Classical Molecular Dynamics (MD) Setup'') was conducted at 300\,K, 
while the top wall was gradually lowered by 0.1~\AA\ every 100 steps until the density 
reached 1.0\,g/cm$^3$.

\begin{figure}[htbp]
  \centering
  \includegraphics[width=0.8\linewidth]{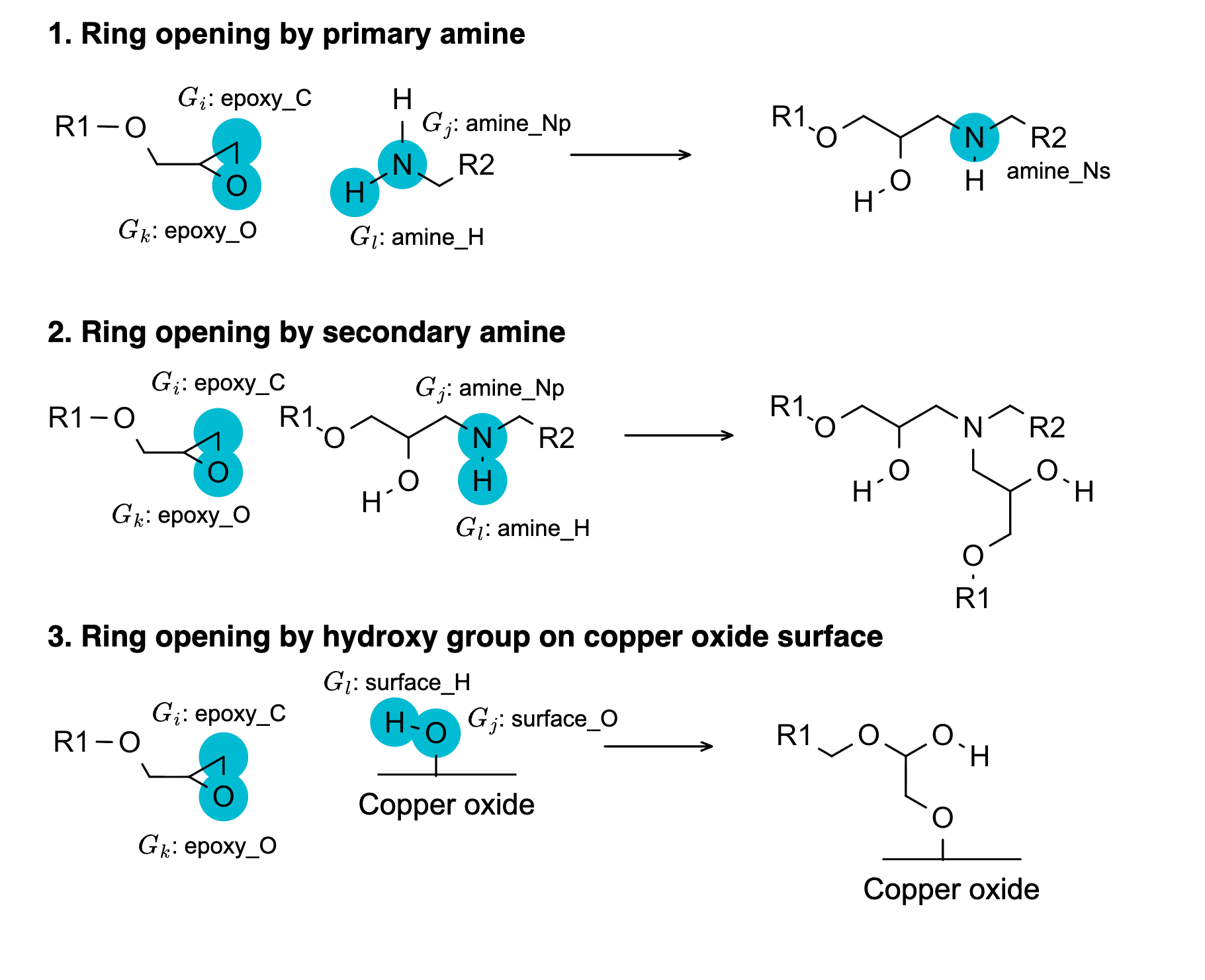}
  \caption{
Reaction patterns for epoxy curing: (1) ring opening by a primary amine, (2) ring opening by a secondary amine, and (3) ring opening initiated by a hydroxyl group on CuO(001).
Reactive centers subject to bias are highlighted.
}
  \label{fig:reaction_epoxy}
\end{figure}

The equilibrated epoxy/amine mixture was then placed on top of the hydroxylated CuO slab 
to assemble the interface model. 
Subsequent equilibration was performed in the {\it NVT} ensemble at 300\,K for 25\,ps using PFP. 
Production simulations using reactive acceleration MD were then performed in the {\it NVT} ensemble 
at 333\,K. 
Temperature and pressure were controlled using a Langevin thermostat. 
The reaction patterns included in the acceleration scheme are summarized in 
Fig.~\ref{fig:reaction_epoxy}, and the corresponding distance criteria and bias assignments 
are listed in Table~\ref{tab:reaction_params_epoxy}.

\vspace{2ex}

\begin{table}[h]
\centering
\caption{Search parameters and bias assignment for epoxy curing at the CuO interface. Distances in \AA.}
\label{tab:reaction_params_epoxy}
\begin{tabular}{lcccccccccc}
\toprule
\multirow{4}{*}{Reaction} & \multicolumn{6}{c}{Group identification} & \multicolumn{2}{c}{Applied bias} \\
\cmidrule(lr){2-7} \cmidrule(lr){8-9}
 & \multicolumn{2}{c}{$i$–$j$} & \multicolumn{2}{c}{$i$–$k$} & \multicolumn{2}{c}{$j$–$l$} & $V^{f}$ & $V^{d}$ \\
\cmidrule(lr){2-3} \cmidrule(lr){4-5} \cmidrule(lr){6-7}
 & $r_{\min}$ & $r_{\max}$ & $r_{\min}$ & $r_{\max}$ & $r_{\min}$ & $r_{\max}$ & & \\
\midrule
RO by primary amine   & 3.0 & 6.0 & 0.0 & 3.0 & 0.0 & 3.0 &  $i$–$j$, $k$–$l$     & $i$–$k$, $j$–$l$ \\
RO by secondary amine   & 3.0 & 6.0 & 0.0 & 3.0 & 0.0 & 3.0 &  $i$–$j$, $k$–$l$     & $i$–$k$, $j$–$l$ \\
RO by hydroxyl group   & 3.0 & 6.0 & 0.0 & 3.0 & 0.0 & 3.0 &  $i$–$j$, $k$–$l$     & $i$–$k$, $j$–$l$ \\

\bottomrule
\end{tabular}
\end{table}

To ensure statistical reliability, three independent simulations were performed with randomized 
initial configurations, and the reported results represent averages over these runs.

\vspace{2ex}

\section{Boost Potential Parameter Dependency}

\begin{figure}[htbp]
  \centering
  \includegraphics[width=\linewidth]{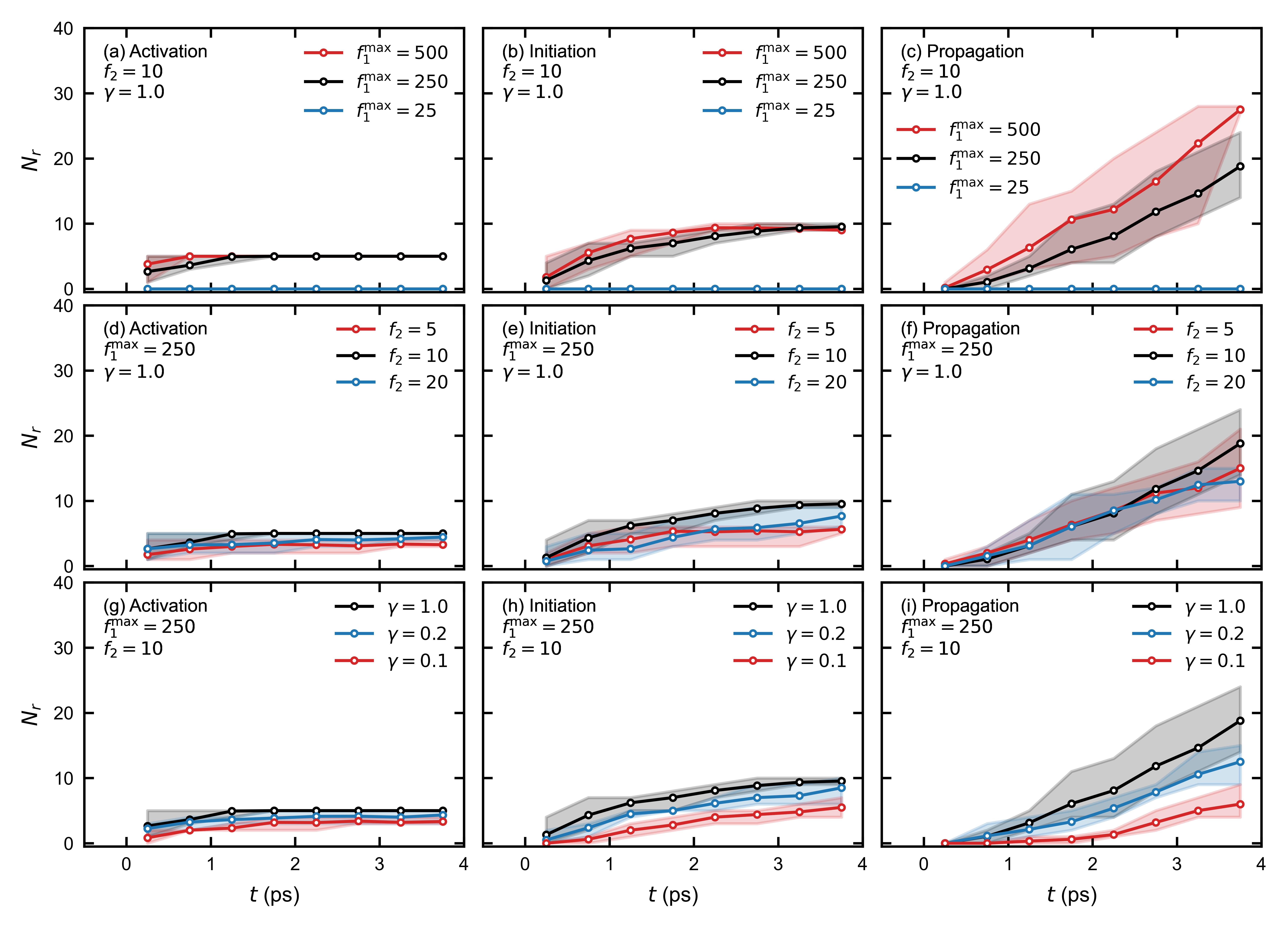}
  \caption{
Parameter sensitivity analysis of the time-dependent bond-boost scheme 
using styrene radical polymerization as a representative system. 
Panels (a--c) show the effect of varying $f_{1}^{\mathrm{max}}$ 
(25, 250, 500 kcal/mol) on (a) activation, (b) initiation, and (c) propagation events. 
Panels (d--f) illustrate the effect of varying $f_{2}$ (5, 10, 20), 
while panels (g--i) show the effect of $\gamma$ (0.1, 0.2, 1.0). 
Each curve represents the cumulative number of reactions ($N_r$) as a function of simulation time, 
with shaded regions indicating the standard deviation across three independent runs. 
The results demonstrate that $f_{1}^{\mathrm{max}}$ exerts limited influence 
once it exceeds a moderate threshold, $f_{2}$ variations yield negligible changes, 
and $\gamma$ acts primarily as a global scaling factor for reaction rates. 
These findings confirm that the default parameter set 
($f_{1}^{\mathrm{max}}=250$/$125$ kcal/mol, $f_{2}=10$, $\gamma=1.0$) 
is broadly representative and does not distort relative reactivity trends.
}
  \label{fig:parameter_sensitivity}
\end{figure}

To assess the robustness of the time-dependent bond boost parameters, 
we systematically varied $f_{1}^{\mathrm{max}}$, $f_{2}$, and $\gamma$ 
using styrene radical polymerization as a representative system. 
The cumulative number of reactions was evaluated as a function of simulation time 
for each parameter set (Fig.~\ref{fig:parameter_sensitivity}).

For $f_{1}^{\mathrm{max}}$, values below $\sim$25\,kcal/mol 
suppressed reactions within the specified biased time window, 
whereas values above $\sim$250\,kcal/mol yielded nearly identical behavior. 
This insensitivity arises because once the bias is sufficiently strong to trigger 
bond formation within the time window, the exact timing becomes irrelevant: 
The reactive-pair list is reinitialized after each event, rendering the magnitude 
of $f_{1}^{\mathrm{max}}$ unimportant beyond the threshold.

For $f_{2}$, the system behavior remained essentially unchanged across the tested range (5--20). 
Although $f_{2}$ formally controls the spatial extent of the bias, 
polymerization is governed by near-contact events, 
and $f_{2}\approx 10$ was found to provide robust performance.

In contrast, varying $\gamma$ directly affected the global rate of reaction. 
Larger $\gamma$ values uniformly accelerated all reaction channels, 
effectively acting as a scaling factor for the overall kinetics. 
Importantly, no evidence of pathway-specific bias was observed.

These results confirm that the default parameter set 
($f_{1}^{\mathrm{max}}=250$\,kcal/mol for $V^f$, 
$f_{1}^{\mathrm{max}}=125$\,kcal/mol for $V^d$, 
$f_{2}=10$, and $\gamma=1.0$) is broadly representative 
and does not distort relative reactivity trends.

\section{Activation energy}

Activation energy estimates were obtained for representative propagation reactions. 
As a model system, we selected an AIBN-derived monomer radical (propagation chain end) 
and a monomer unit. 
For each monomer, an initial product structure was first constructed, in which 
the radical had reacted with the $\beta$-carbon of the monomer. 
To generate an approximate transition-state geometry, the distance between 
the radical center and $\beta$-carbon was incrementally varied, 
yielding a set of initial guesses along the reaction coordinate. 

These guesses were refined using the ReactionString method provided by Matlantis~\cite{Matlantis} to obtain
optimized transition-state structures. 
For each case, the energies of the reactant, transition state, 
and product were evaluated. 
The procedure was repeated for ten distinct conformations of each monomer, 
and the lowest-energy conformers for both the reactant and product 
were used to define the activation energy as the difference between 
the transition state and lower of the two endpoint energies. 

This protocol ensured that conformational sampling effects were accounted for and provided a consistent measure of the propagation barriers used in the comparison with simulation-derived rate constants 
(Table~\ref{tab:kp_Ea_rho}).

\begin{table}[htbp]
  \centering
  \small
  \begin{threeparttable}
    \caption{Comparison of polymerization parameters from PFP with time-dependent bond-boost (TDBB) simulated values and experimental data.}
    \label{tab:kp_Ea_rho}
    \begin{tabular}{
      l
      S[table-format=5.0] S[table-format=2.2] S[table-format=1.2]
      S[table-format=5.0] S[table-format=2.2] S[table-format=1.2]
    }
      \toprule
      {Monomer \tmark{a}} &
      \multicolumn{1}{c}{TDBB} &
      \multicolumn{2}{c}{PFP} &
      \multicolumn{3}{c}{Experiment} \\
      \cmidrule(lr){2-2}\cmidrule(lr){3-4}\cmidrule(lr){5-7}
      & {$k_\mathrm{p}^*$ (333 K)} & {$E_\mathrm{A}$ (kcal/mol)} & {$\rho$ (g/cm$^3$)}
      & {$k_\mathrm{p}$ (333 K)}& {$E_\mathrm{A}$ (kcal/mol)} & {$\rho$ (g/cm$^3$)} \\
      \midrule
      MA   & {2.62} & {5.5} & {0.97} & {31400\cite{tsarevsky2013fundamentals}} & {4.4\cite{tsarevsky2013fundamentals}} & {0.95\cite{haynes2016crc}} \\
      VAc  & {2.14} & {6.3} & {0.96} & {2300\cite{odian2004principles}}  & {4.3\cite{odian2004principles}} & {0.93\cite{haynes2016crc}} \\
      MMA  & {1.56} & {6.0} & {0.96} & {515\cite{odian2004principles}}   & {6.3\cite{odian2004principles}} & {0.93\cite{haynes2016crc}} \\
      St   & {2.68} & {5.0} & {0.95} & {165\cite{odian2004principles}}   & {6.2\cite{odian2004principles}} & {0.90\cite{haynes2016crc}} \\
      ICDM & {1.38} & {8.5} & {1.10} & {27\cite{tsarevsky2013fundamentals}} & {6.0\cite{tsarevsky2013fundamentals}} & {1.12\cite{haynes2016crc}}\\
      DPE  & {0.78} & {15.3} & {1.05} & {- \tmark{a}} & {- \tmark{a}} & {1.02\cite{sigmaaldrich-d206806}} \\
      \bottomrule
    \end{tabular}
    \begin{tablenotes}
      \footnotesize
      \item[a] MA: methyl acrylate, MMA: methyl methacrylate, St: styrene, VAc: vinyl acetate, 
ICDM: dimethyl itaconate, DPE: diphenylethylene,
      \item[b] DPE generally does not polymerize under standard radical polymerization conditions, although several studies have reported polymerization under specific conditions.\cite{sato2001effect,zhang2023intramolecular}
    \end{tablenotes}
  \end{threeparttable}
\end{table}
\bibliography{references}